# Martensitic transformation induced by cooling NiTi wire under various tensile stresses: martensite variant microstructure, textures, recoverable strains and plastic strains

O. Tyc, X. Bian, O. Molnárová, L. Kadeřávek, L. Heller, P. Šittner

Institute of Physics of the Czech Academy of Sciences, Na Slovance 2, Prague, Czech Republic

**Abstract**

To understand how martensitic transformation (MT) in polycrystalline NiTi shape memory alloy (SMA) proceeds under external stress, we evaluated recoverable transformation strains and plastic strains generated by the forward MT in nanocrystalline NiTi wire cooled under tensile stresses 0-600 MPa, determined textures in martensite and reconstructed martensite variant microstructures within the selected grains of the cooled wire.

The obtained findings show that the forward MT proceeding under external stresses gives rise to characteristic recoverable transformation strains, plastic strains, martensite variant microstructures and martensite textures. MT occurring upon cooling under stresses exceeding 100 MPa creates martensite variant microstructures consisting of single domain of partially detwinned laminate of (001) compound twins filling entire grains. The recoverable transformation strain increases with increasing stress, reaches maximum ~5% at ~200 MPa stress, and remains constant with further increasing stress up to 600 MPa. Starting from 100 MPa stress, the forward MT generates also plastic strains, the magnitude of which also increases with increasing stress. The texture of martensite in the wire cooled under increasing external stress gradually changes from four (8)-pole texture of the self-accommodated martensite towards single-pole (10-3) fiber texture of oriented martensite forming under 200 MPa stress. Martensite variant microstructures and martensite textures observed in NiTi wires cooled under high stresses ~600 MPa are yet different evidencing plastic deformation of martensite.

We propose that, in the absence of external stress, the forward MT takes place via propagation of strain compatible habit plane interfaces between austenite and second order laminate of (001) compound twins. When the forward MT takes place under external tensile stress, it occurs equally, but the newly created martensite immediately reorients into single domains of (001) compound twins, partially detwins and deforms plastically under the action of the external stress. The plastic strain generated by the forward MT upon cooling under stress is attributed to the [100](001) dislocation slip in martensite at low stresses and plastic deformation of martensite by kwinking at high stresses.

# 1. Introduction

When NiTi shape memory alloy polycrystalline wire in austenitic state is cooled and/or deformed mechanically, martensitic transformation from the B2 cubic austenite to B19' monoclinic martensite creates martensite variant microstructures, martensite textures, generates recoverable transformation strains and potentially unrecoverable plastic strains. During the martensitic transformation, blocks of mutually misoriented martensite variants (each providing large strain gradient with respect to the parent austenite) arrange in a self-accommodated manner within martensite variant microstructures in such a way that the material sample does not change its external shape.

Martensite variant microstructures created by the B2 to B19' martensitic transformation upon stress-free cooling were investigated by transmission electron microscopy (TEM) in the last decades [1-7]. The experimentally observed martensite variant microstructures forming self-accommodated martensite were frequently found to consist of <011> type-II twin or (111) type-I twin laminates arranged in a triangular geometry in the case of NiTi single crystals or coarse grained NiTi polycrystals [1-5]. Aleternatively, multiple domains of (001) compound twin laminate were observed in nanocrystalline NiTi alloys [6,7]. Twin laminates appear in the martensite variant microstructure created with or without stress. The laminates form either because the martensite nucleated as a twinned structure and/or because the twins assure strain compatibility at habit plane interfaces (the twinning modes act as lattice invariant shear (LIS) in the newly formed martensite phase) in accord with predictions by the Phenomenological Theory of Martensite Crystallography (PTMC) [8,9]. Based on the latter, the laminated martensite domains are called habit plane variants (HPV). MT taking place via strain compatible habit plane conforming the PTMC theory shall not generate any permanent lattice defects. In spite of that, some investigators reported that MT proceeding in the absence of stress, generates dislocation defects [3,4]. It depends on the material state of the investigated NiTi alloy, while MT upon thermal cycling of solution annealed NiTi alloys generate dislocation defects [3,4], no dislocation defects were observed in thermally cycled nanocrystalline NiTi alloys [10].

Akamine et al. [3] recently investigated self-accommodated martensite variant microstructures and lattice defects created by forward MT thermally induced during the stress-free cooling of solution annealed NiTi by TEM. They observed martensite variant microstructures consisting of <011> type-II twins in martensite and dislocation loops in austenite created presumably during the reverse MT at the intersection between the martensite twins and habit plane. They concluded that the observed dislocations were nucleated at the steps in habit plane formed by the martensite twins during the reverse MT. No lattice defects were assumed to be created during the forward MT on cooling. This suggests significant asymmetry between the forward and reverse MT. Nevertheless, since no external stress was applied and since strain is assumed to be compatible at the propagating habit plane interfaces, this result cannot be generalized.

From the application point of view, however, more important are martensite variant microstructures created by the martensitic transformation taking place under external stress - e.g. upon isothermal tensile loading or upon cooling under external stress. Such martensitic transformation creates martensite variant microstructures containing preferentially oriented martensite variants, gives rise to characteristic martensite textures, reversible shape changes and potentially plastic strains in closed loop thermomechanical tests. Whether the MTs proceeds without or under stress, it takes place via propagation of habit plane interfaces, which, together with the applied stress state, critically affects the formation of martensite variant microstructures in grains of nanocrystalline NiTi wires. While habit planes propagating during the thermally induced MT in NiTi were thoroughly investigated in the literature [1-6], there is no general agreement on the habit planes of the MT proceeding under external stress.

The martensitic transformation proceeding under external stress in NiTi polycrystals generates recoverable transformation strains, plastic strains [10-16] and permanent lattice defects [3,4,10,12,15,16,17]. The plastic strains are generated repeatedly during thermomechanical cycling, whenever the forward or reverse MT take place above certain stress thresholds [10,18]. In closed loop thermomechanical load cycles, this gives rise to accumulation of residual plastic strains [12,13], internal stress [19] and lattice defects [10,12,14,16], which causes instability of cyclic thermomechanical responses called "functional fatigue" [20]. Although the functional fatigue has been investigated theoretically and experimentally for decades, the mechanism by which stress induced B2-B19' martensitic transformation in NiTi generates plastic strains and permanent lattice defects remains blurred.

Researchers, who studied the stress induced martensitic transformations in NiTi [21-24], focused naturally on the generation of dislocation defects at the habit plane propagating during the forward MT on loading because the stress is higher on loading. They assumed stress induced martensitic transformation into <011> type II twinned martensite [25] predicted by the Phenomenological Theory of Martensite Crystallography (PTMC) [8]. The idea proposed in [16,21-25] is that the strain compatible habit plane interfaces propagating during the MT nucleate slip dislocations at its crystallographic defects (steps, disconnections). This mechanism may possibly work for NiTi alloys which display <011> type II twinned martensite [3.4,21,22,23] during the reverse transformation on unloading and/or heating (austenite remains behind the moving interface), but not during the forward transformation upon mechanical loading or cooling under stress [26]. On the other hand, this mechanism is not applicable to NiTi alloys which display (001) compound twins in martensite variant microstructures [6,7] simply because there are no <011> type II twins in the martensite variant microstructure created by the forward MT.

Generally, there is a problem with the lack of experimental evidence on martensite variant microstructures and permanent lattice defects created by the forward MT taking place under external stress. It remains unclear how strain compatibility at the habit plane is achieved during the forward MT in NiTi proceeding into detwinned or (001) compound twinned martensite under external stress. It is also unclear up to which extent is the strain compatibility at the habit plane responsible for plastic strain generated by the MT under stress.

Strain compatibility at the habit plane interfaces propagating under external stress was also investigated theoretically. While the PTMC theory considers transformation strains only, analysis of the strain compatibility at the habit plane interface propagating under external stress needs to consider total strains due to elastic, transformation and plastic deformation of the crystal lattices on both sides of the interface. Xiao et al. [27] assumed that strain compatibility at the habit plane interface between austenite and detwinned martensite can be achieved by considering plastic deformation via dislocation slip in austenite (PTMC theory updated using local plastic strains due to dislocation slip in austenite [28]). Heller and Sittner [29] studied the effect of elastic deformation of both austenite and martensite lattices due to external stress on the formation of strain compatible habit plane interfaces. They proposed that habit plane interfaces can form between austenite and detwinned martensite on crystal planes, the indices of which are not firmly given by lattice parameters and lattice correspondence between austenite and martensite, but vary with the magnitude and orientation of the applied stress. They considered softening of austenite elastic constants on cooling but neglected possible softening of martensite elastic constant on heating as well as plastic deformation. Evidently, more experimental and theoretical research on habit planes and martensite variant microstructures created by the forward MT proceeding under external stress is needed.

Motivated by the need to evaluate martensite variant microstructures and martensite textures created by MT under external stress, we have recently developed two experimental methods for: i) post mortem reconstruction of martensite variant microstructures in deformed NiTi wires [30] and ii) in-situ analysis of austenite and martensite textures evolving during tensile test on NiTi shape memory wires up to the fracture [31]. These methods proved to be instrumental for the discovery of the mechanism of plastic deformation of B19' martensite via coupled [100](001) dislocation slip based kinking and (100) twinning in martensite called kwinking [32,33]. Kwink interfaces are internal interfaces in plastically deformed martensite created by plastic slip and twinning [33].

In this work, we have applied these two methods to the analysis of the forward MT upon cooling under external stress in nanocrystalline NiTi SME wire. We evaluated recoverable transformation and plastic strains, reconstructed martensite variant microstructures and determined martensite textures created by the forward MT upon cooling the wire under wide range of external stresses 20-600 MPa. Martensite variant microstructures, textures and plastic strains created by the forward stress induced MT in tensile tests at various constant temperatures on the same NiTi SME wire are reported in closely related work [14].

## 2. Experiment

NiTi shape memory wire produced by Fort Wayne Metals in cold work state (FWM #5 Ti-50.5 at. % Ni, 42 % CW, diameter 0.1 mm) was heat treated by short pulse of electric current [34] (power density 160 W/mm$^3$, pulse time 15 ms). While performing the heat treatment, 30 mm long segment of cold worked wire was crimped by two steel capillaries, prestressed to ~300 MPa, constrained in length and subjected to 15 ms pulse of controlled electric power. The heat treated wire has a fully recrystallized microstructure with a mean grain size d = 250 nm. The alloy undergoes B2-B19' transformation with characteristic transformation temperatures $M_s$ = 63 °C, $A_f$ = 93 °C upon cooling/heating, as determined by combined in-situ electric resistivity and dilatometry test (Fig. 1).

The 15 ms NiTi #5 wire was selected because: i) it is martensitic at room temperature, ii) forward martensitic transformation is accompanied by generation of significant plastic strains and iii) martensite variant microstructures in 250 nm diameter grains can be reconstructed by TEM. Reconstruction of martensite variant microstructures in grains of the nanocrystalline NiTi wire by TEM requires the grains to be large enough so that they do not overlap in the 50 nm thin TEM lamella cut from the deformed wire, but small enough so that the martensite variant microstructure within whole grain can be analyzed. Moreover, grains in virgin wire samples prior to tests have to be clear from lattice defects persisting from the cold work to enable analysis of lattice defects created by the tensile deformation. Thermomechanically induced MTs in selected 15 ms NiTi #5 shape memory wire were already investigated in Refs. [35,43].

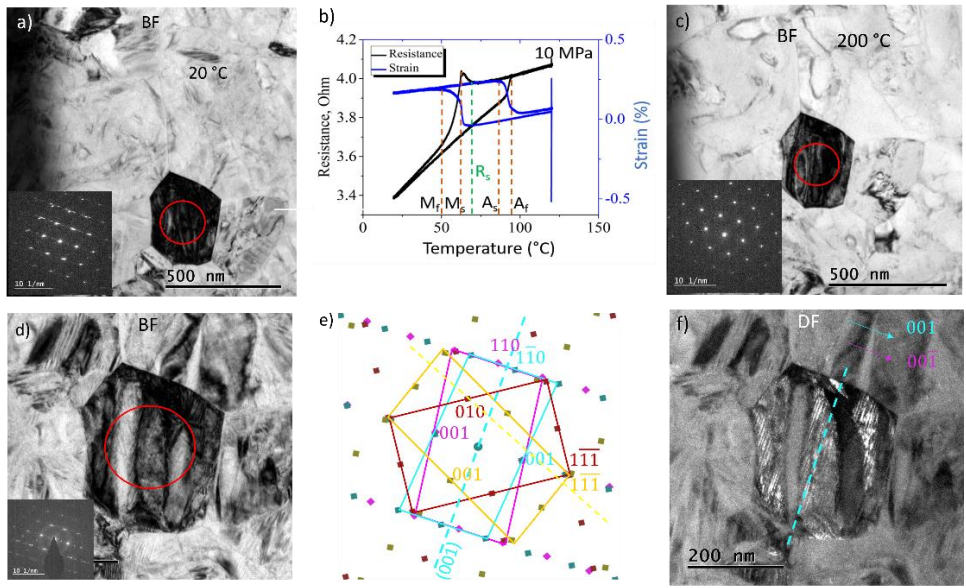

**Fig. 1: Thermally induced B2-B19' martensitic transformation in 15 ms NiTi #5 shape memory wire.** Transformation temperatures evaluated by combination of dilatometry and electric resistivity measurement. TEM analysis shows ~ fully recrystallized austenitic microstructure (grain size ~250 nm) and self-accommodated martensite variant microstructure consisting of multiple (001) compound twin laminates with grains.

Tensile tests were carried out using an in-house-built tensile tester (MITTER) consisting of a miniature load frame, an environmental chamber, electrically conductive grips, a load cell, a linear actuator and a position sensor. The environmental chamber (Peltier elements) enables to maintain homogeneous temperature around the thin wire from -30 °C to 200 °C. The wire sample was gripped into the tester, strain was set to zero in the austenitic state at 200 °C, the wire was cooled/heated under 20 MPa stress to the desired test temperature and tensile thermomechanical loading tests were performed. Thermomechanical loading tests were performed in combined position and force control mode (cooling rate of 3 °C/min). Thermomechanical tests involving cooling down to −120 °C and above 200 °C were performed using DMA 850 tester by TA Instruments using 10 mm long wire samples. Thermal strains were calibrated using quartz sample and subtracted from the signal recorded in thermomechanical tests. The wire sample was gripped into the tester, strain was set to zero in the austenitic state at high temperature above $A_f$, and thermomechanical loading tests was run in combined position and load control mode as described in Fig. 2c,d.

Thin lamellae for TEM analysis were cut from the subsurface layers of deformed wire (10 μm below the surface, wire axis in the lamella plane) by focused ion beam (FIB) using a FEI Quanta 3D FIB-SEM microscope. Martensite variant microstructures in grains of deformed NiTi wire were analysed by TEM and HRTEM methods using a FEI Tecnai TF20 X-twin microscope equipped with a field emission gun operating at 200 keV using a double tilt specimen holder. The recorded electron diffraction patterns were indexed using the lattice parameters and the atomic positions in the B19' structure (Table 1).

**Table 1:** Lattice parameters and atomic positions in unit cell of B2 cubic austenite and B19' monoclinic martensite in NiTi [25,37] used in indexing the electron diffraction patterns.

| Cell | Lattice parameters [nm] | | | | Atomic positions in unit cell [nm] | | | | Cell volume [nm³] |
|------|-----|-----|-----|-----|-----|-----|-----|-----|-----|
|      | a | b | c | β | Ti1 | Ti2 | Ni1 | Ni2 | |
| B2   | 0.3015 | | | | 0.0000 | | 0.5 | | 0.0274 |
|      | | | | | 0.0000 | | 0.5 | | |
|      | | | | | 0.0000 | | 0.5 | | |
| B19' | 0.2889 | 0.4120 | 0.4622 | 96.80 | 0.0000 | 0.1220 | 0.5950 | 0.5270 | 0.0546 |
|      | | | | | 0.0000 | 0.5000 | 0.0000 | 0.5000 | |
|      | | | | | 0.0000 | 0.5340 | 0.4370 | 0.0970 | |

It was assumed that martensite variant microstructures, which existed in the NiTi wire cooled under tensile stress to the room temperature were retained in the TEM lamella cut from the deformed wire. Martensite variant microstructures were analysed by Selected Area Electron Diffraction with Dark Field (SAED-DF) method (see Appendix A in [7]) and by the automated orientation mapping ASTAR method [36].

Reconstruction of the martensite variant microstructure by the ASTAR method is achieved by scanning the selected grain by precessing electron beam while recording electron diffraction patterns and images by fast CCD camera [36]. For more information on the ASTAR reconstruction of martensite variant microstructure in NiTi see Ref. [30].

When using both SAED-DF and ASTAR methods to analyse the martensite variant microstructures in whole grain, TEM lamella is tilted in such a way that the selected grain is oriented into <010> low index zone (<010> denotes [010] and/or [0-10] direction of the monoclinic lattice). The <010> zone is selected because the martensite variants in grains arrange themselves in such a way that all crystal lattices within a single grain are equally oriented with respect to the electron beam in the [010] zone and all interface planes are parallel to the electron beam [30]. This is direct consequence of the activated deformation mechanisms, particularly to martensite reorientation and plastic deformation of martensite by kwinking [30,32,33].

In situ synchrotron x-ray diffraction experiment were conducted at the ID15A beamline at ESRF Grenoble, using 64.2 keV beam energy (wavelength 0.19312 Å) and a beam size of 150 × 150 μm$^2$. The MITTER wire tester was installed on the beamline so that the gauge volume was placed into the wire center. The wire samples were subjected to closed loop thermomechanical loading tests (Fig. 3) and 2D diffraction patterns were recorded continuously during tests taking advantage of the high intensity of the diffracted beams and low exposure time 0.2489 s of the used Pilatus3 X CdTe 2M detector.

Data analysis was performed as follows. 2D diffraction images were regrouped into 2048 bins along radial (2$\theta$) coordinates with 2$\theta$ spacing from 0.34562° to 18.12745° to include main reflections. All ring patterns were segmented into Δη=2° slices such that each recorded 2D diffraction pattern yields 111 radially integrated 1D diffraction spectra of 2$\theta$ versus total x-ray intensity. Thus, each 1D diffraction spectrum provides orientation specific information on the microstructure within the gauge volume within ±1° angle around the corresponding azimuth angle. The set of 1D diffraction spectra from the CeO$_2$ standard was used to determine the azimuth-dependent instrumental contribution to profile line broadening.

To evaluate austenite and martensite textures from x-ray diffraction patterns, Rietveld refinement of the whole set of radially integrated 1D diffraction spektra (111 in total) obtained by sequencing Debye-Scherrer ring patterns was performed using the General Structure and Analysis System (GSAS-II) [38] with a modified script from GSASII scriptable routines [39]. A generalized spherical-harmonic description of texture was implemented [40]. It is known that intensity variation along a Debye-Scherrer ring (i.e., corresponding to a *{hkl}* reflection) is in proportion to pole intensity of the orientation sphere imposed on a sample. Given that no sample rotation was applied and viewing along the AD direction, each Debye-Scherrer ring results in pole figure coverage, which is represented as a pair of centrosymmetric curved lines with an interspacing of 2$\theta$ on the *{hkl}* pole figure [41]. Texture was processed in the program MTex [42].

# 3. Results

## 3.1 Recoverable and plastic strains generated by forward MT upon cooling under external stress

When NiTi wire is subjected to thermal cooling-heating cycle under external tensile stress (Fig. 2a,b), it elongates upon cooling due to the forward MT and shortens upon heating due to the reverse MT. If plastic strain is generated by the forward and/or reverse MT during the closed loop thermomechanical load cycle, the parent austenite phase is restored but the original length of the wire is not recovered completely (Fig. 2).

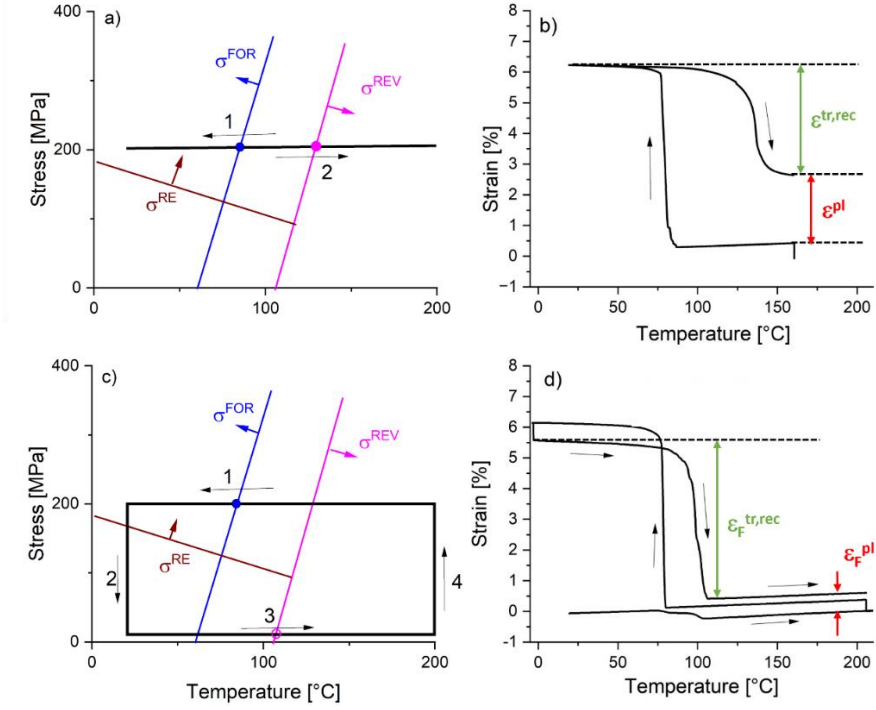

**Fig. 2: Thermomechanical loading tests on 15 ms NiTi#5 shape memory wire to determine recoverable transformation strain and plastic strains generated by the martensitic transformation under stress** in: a,b) standard thermal cycle under constant stress and c,d) thermomechanical loading tests along rectangle stress-temperature path in which plastic strains are generated by the single forward MT proceeding under external stress. The [temperature, stress] conditions, at which the forward and reverse MT occurred under external stress are denoted by solid circles (a,c). See Supplementary materials for further details.

The recoverable transformation strain $\varepsilon^{tr,rec}$ and plastic strains $\varepsilon^{pl}$ strains evaluated as shown in Fig. 2a,b are, however, neither recoverable transformation strain nor the plastic strain generated by the forward MT. This is not a problem for engineers who investigate actuation performance of NiTi wires, since the $\varepsilon^{tr,rec}$ and $\varepsilon^{pl}$ strains are perfectly relevant for evaluation cyclic stability of actuation in applied research. However, if one wants to reveal the mechanism by which the forward MT proceeding under stress generates plastic strain, true recoverable transformation strain $\varepsilon_F^{tr,rec}$ and plastic strains $\varepsilon_F^{pl}$ strains generated by the forward MT have to be evaluated as shown in Fig. 2c,d. Clearly, the $\varepsilon^{tr,rec}$ and $\varepsilon_F^{tr,rec}$ strains as well as $\varepsilon^{pl}$ and $\varepsilon_F^{pl}$ strains are significantly different.

The true recoverable transformation strain $\varepsilon_F^{tr,rec}$ and plastic strains $\varepsilon_F^{pl}$ strains generated by the forward MT can be evaluated from the closed loop test performed along rectangular stress-temperature path denoted in the σ-T diagram in Fig. 2c because the reverse MT upon heating under 20 MPa does not generate any plastic strain [18,10]. The recoverable transformation strain $\varepsilon_F^{tr,rec}$ and plastic strains $\varepsilon_F^{pl}$ recorded in this test (Fig. 2d) are thus generated by the forward MT only. The plastic strains which are generated by the reverse MT under stress (Fig. 2a,b) affect the $\varepsilon^{tr,rec}$ and $\varepsilon_F^{pl}$ recorded in the standard thermal cycle ($\varepsilon_F^{tr,rec} \gg \varepsilon^{tr,rec}$ and $\varepsilon_F^{pl} \ll \varepsilon^{pl}$). Another very important difference is that the recoverable transformation strain $\varepsilon_F^{tr,rec}$ and plastic strain $\varepsilon_F^{pl}$ generated by the forward MT do not depend on elastic strains of austenite and martensite.

The recoverable transformation strain is of course generated only during the forward MT. The $\varepsilon_F^{tr,rec}$ has a very good meaning, as it corresponds to martensite variant microstructures and martensite texture created by the forward MT under stress and does not depend on elastic strains of austenite and martensite. Plastic strains $\varepsilon^{pl}$, on the other hand, can be generated by both forward and reverse MTs taking place under stress. Plastic strains $\varepsilon^{pl}$ are approximately equal to the sum of plastic strains $\varepsilon_F^{pl}$ and $\varepsilon_R^{pl}$ generated by the forward MT and reverse MT, respectively [11]. As the ultimate goal of this research is to maximize recoverable transformation strains and minimize plastic strains generated by forward and reverse MTs under stress in closed loop thermomechanical cycles, we need to evaluate plastic strains generated separately by forward and reverse MT. In this work, we focus solely on the forward MT on cooling NiTi shape memory wire under various external stresses (Figs. 2,3,4).

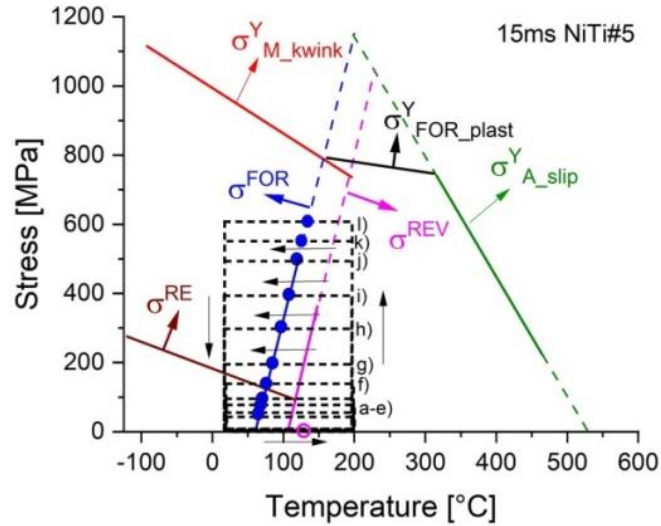

**Fig. 3: Stress-temperature paths of thermomechanical loading tests on 15 ms NiTi #5 SME wire in the σ-T diagram.** The diagram was determined from series of isothermal and isostress tensile tests until rupture in Ref. [43]. The thermomechanical loading tests were performed along rectangular stress-temperature paths to evaluate recoverable transformation strains, plastic strains, martensite variant microstructures and textures generated by forward MT upon cooling under various constant stresses and temperatures (temperatures and stresses at which the forward MTs take place are marked by solid blue circles).

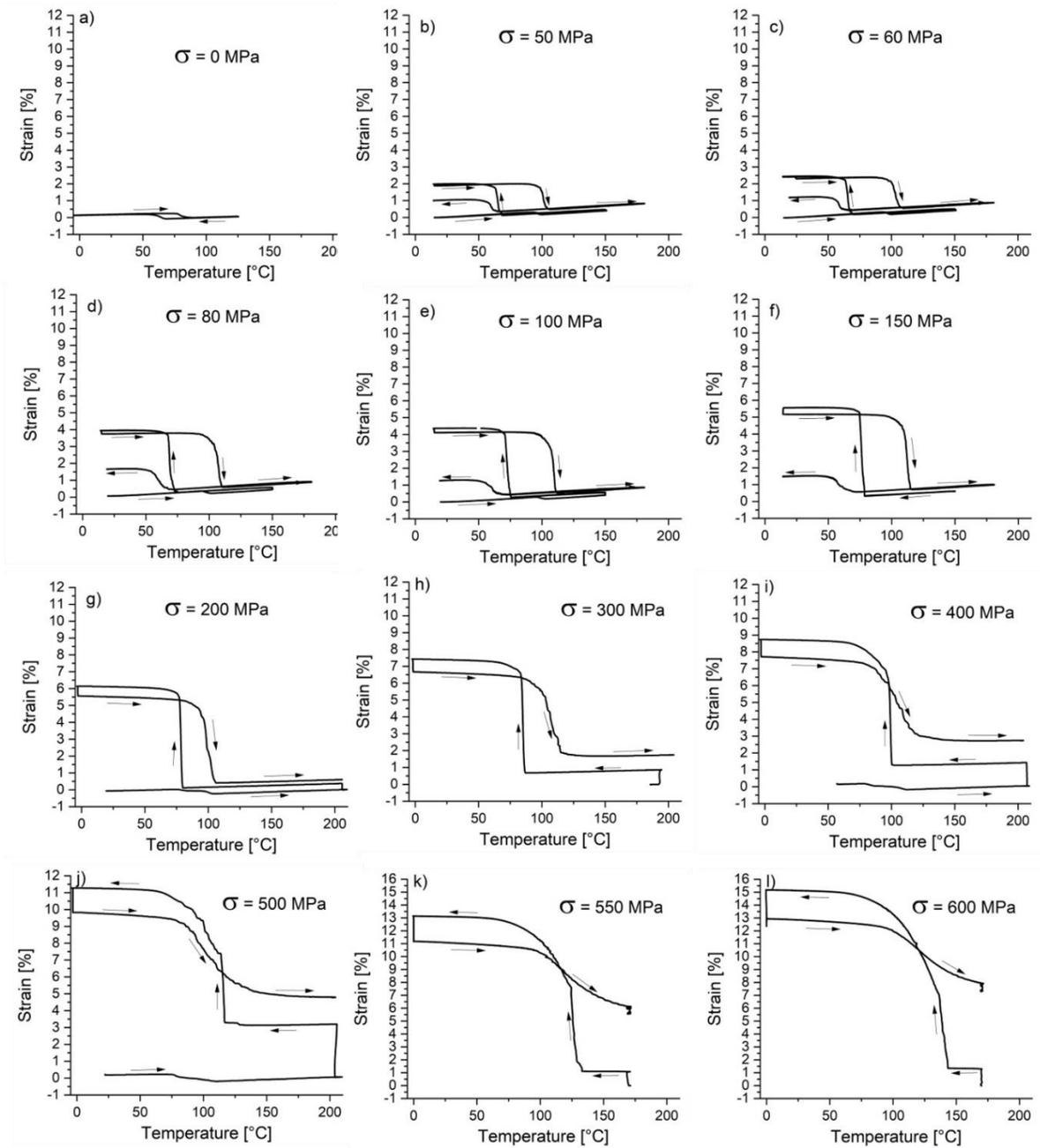

**Fig. 4: Strain – temperature responses evaluated in thermomechanical loading tests on the 15 ms NiTi #5 SME wire** along stress-temperature paths in Fig. 3. The tests in (a-e) start by heating the wire under 20 MPa stress up to 200 °C and finish by cooling under 20 MPa stress to evaluate TWSME effect introduced by cooling under stress. In other tests (f-l), the supplemental starting and/or finishing branches are missing.

The methodology used to determine plastic strains generated separately by the forward and reverse MTs was originally proposed by Heller et al. [18]. It is based on two key assumptions. First is that the forward and reverse MT proceeding under stress above certain thresholds generate plastic strain, but the forward and reverse MTs proceeding in the absence of external stress do not. Second assumption is that the tensile deformation of martensite at low temperatures up to the yield stress does not generate plastic strains. This second assumption is fulfilled in most of nanocrystalline NiTi wires (Fig. S1), but not in the 15 ms NiTi #5 SME wire studied in this work (Fig. S2). However, since we deal only with plastic strains generated by the forward MT on cooling under stress, the failure of the 15 ms NiTi #5 SME wire to meet this second assumption does not pose a problem.

A series of closed loop thermomechanical loading tests were performed along rectangular stress-temperature paths in stress-temperature diagram of the 15ms NiTi #5 SME wire [14,43] involving cooling under applied stresses 0-600 MPa (Fig. 3). The strain-temperature responses recorded in these tests are presented in Fig. 4. It needs to be pointed out that the rectangular stress-temperature paths prescribed in all tests (Fig. 3) lie below the critical [stress, temperature] conditions for plastic deformation of martensite by kwinking $\sigma^Y_{M\_kwink}$, plastic deformation of austenite by dislocation slip $\sigma^Y_{A\_slip}$ and coupled transformation and plastic deformation $\sigma^Y_{FOR\_plast}$ in the σ-T diagram [14,43]. The austenite and martensite phases in the NiTi wire cooled under external stress thus deform plastically in these tests only when the forward MT proceeds (blue solid circles in Fig. 3).

While the forward MTs taking place upon cooling at stresses 0-300 MPa proceed in narrow temperature ranges (Fig. 4a-h), a temperature plateau followed by the additional strain increase on further cooling is observed in tests at higher stresses. The tensile deformation upon cooling is localized in martensite band fronts propagating along the wire at nearly constant temperatures within these temperature plateaus. We evaluate the temperature in the mid of the plateau as the temperature, at which the forward MT takes place (Fig. 4 i-l). This approximation is equivalent to the consideration of the plateau stress in isothermal tensile test as the stress, at which the stress induced MT proceeds [14]. The martensitic transformation does not have to be completed within the temperature (stress) plateau but may continue upon further cooling (loading) beyond the end of the plateau. Nevertheless, recoverable transformation strains and plastic strains are generated during the temperature (stress) plateau as well as during the further cooling (loading) beyond the end of the plateau [14].

Results of the thermomechanical loading tests from Fig. 4 are shown in Fig. 5 in a form stress and temperature dependence of total strain $\varepsilon^{tot}$, recoverable transformation strain $\varepsilon_F^{tr,rec}$ and plastic strain $\varepsilon_F^{pl}$ generated by the forward MT. The strains $\varepsilon^{tot}$, $\varepsilon_F^{tr,rec}$ and $\varepsilon_F^{pl}$ increase with increasing applied stress (Fig. 5a) as well as with increasing temperature, at which the forward MT occurred (Fig. 5b). The recoverable transformation strain $\varepsilon_F^{tr,rec}$ increases with increasing applied stress in the range 0-200 MPa, but remains approximately constant with further increasing stress (Fig. 5a). The plastic strains $\varepsilon_F^{pl}$ increase with increasing applied stress starting from ~150 MPa, which roughly corresponds to the martensite reorientation stress (150 MPa stress, at which the reorientation $\sigma^{RE}$ line crosses the $\sigma^{FOR}$ line in the σ-T diagram (Fig. 3)).

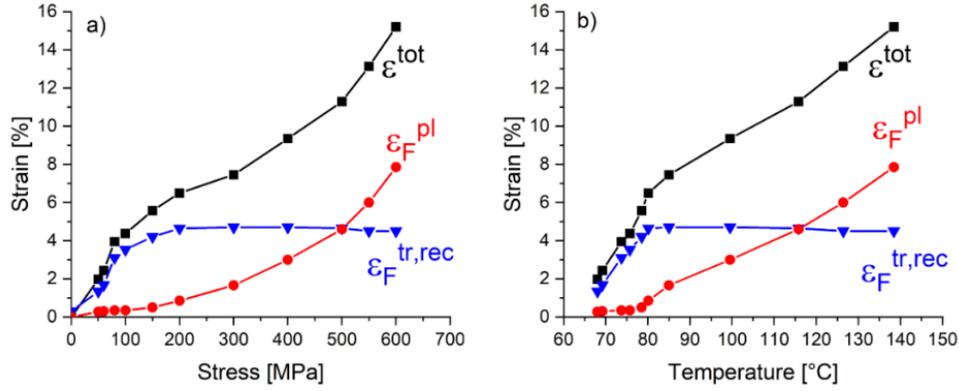

**Fig. 5: Total strain $\varepsilon^{tot}$, recoverable transformation strain $\varepsilon_F^{tr,rec}$ and plastic strain $\varepsilon_F^{pl}$ generated by the forward MT under stress in the range 0-600 MPa** (evaluated from the results of thermomechanical loading tests in Fig. 4) plotted in dependence on the a) external stress and b) temperature, at which the forward MT occurred.

Multiple questions appear based on the experimental results in Figs. 4,5: i) why the maximum recoverable strain is only ~5%, ii) why recoverable transformation strains $\varepsilon_F^{tr,rec}$ do not increase with increasing external stress above 200 MPa, iii) why plastic strains $\varepsilon_F^{pl}$ start to be generated by the forward MT proceeding at 150 MPa, iv) what determines the magnitudes of plastic strains $\varepsilon_F^{pl}$ generated by the forward MT at specific stresses and v) do the plastic strains $\varepsilon_F^{pl}$ apply generally for any kind of thermomechanical loads (e.g. forward MT in isothermal loading). We will address all these questions in section 4.

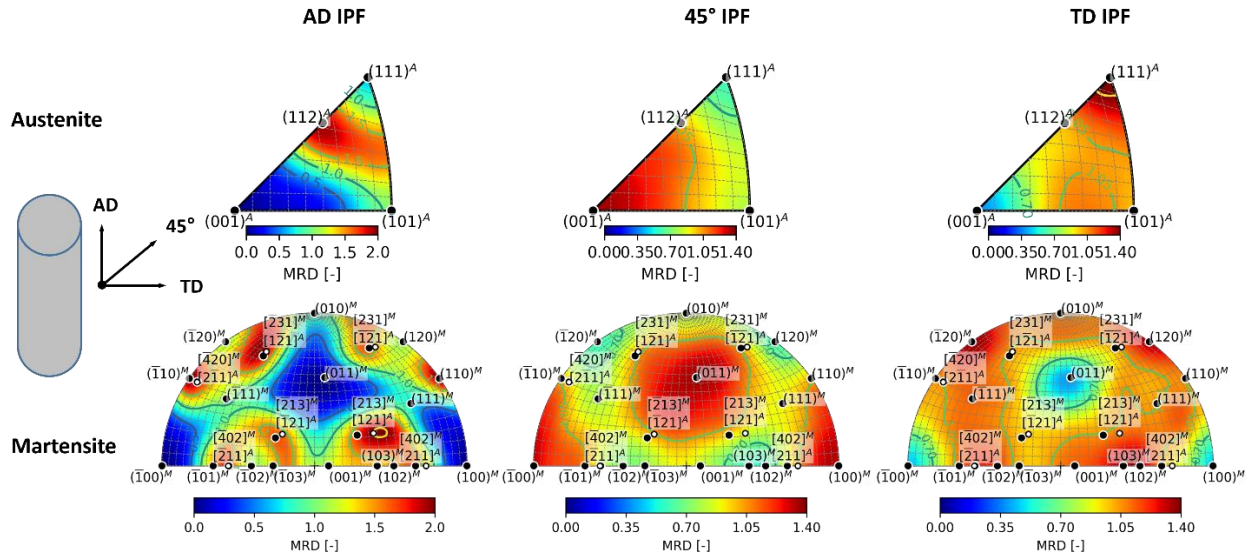

**Fig. 6a: Texture of austenite at 200 °C (top row) and self-accommodated martensite at 20 °C (bottom row) created by cooling 15 ms NiTi #5 wire in the absence of stress.** The textures are presented using inverse pole figures of axial AD, 45° and transverse TD directions to the wire axis. The austenitic fiber texture of NiTi#5 SME wire (AD IPF) displays broad maximum stretching from the [112] pole across the stereographic triangle. This texture transforms into four (8)-pole fiber texture of self-accommodated martensite. The four broad poles in AD IPF actually consist of 8 maxima reflecting the lattice correspondence between <112> austenite and [-420],[-402],[402],[-231],[-213],[213],[231] martensite directions. For more information on martensite fiber textures in NiTi wires see [31].

## 3.2 Textures in martensite created by forward MT upon cooling under external stress

The textures of austenite and martensite textures and volume fraction of phases evolving during thermomechanical loading tests were determined by in-situ synchrotron x-ray experiments. The texture of austenite existing in the NiTi wire at high temperatures 200 °C and texture of self-accommodated martensite existing at room temperature after cooling in the absence of tensile stress are shown in Fig. 6a in a form of inverse pole figures of axial direction (AD IPF), 45° direction (45° IPF) and transverse direction (TD IPF).

The austenite and martensite textures are slightly different from the textures observed in 15 ms NiTi#1 superelastic NiTi wire (Fig. 6b) studied in our closely related experiments [31]. Key difference consists in the fiber texture of the B2 austenite in 15 ms NiTi #5 wire which is less sharp and which displays broad maximum stretching from the <112> pole across the stereographic triangle. During the martensitic transformation upon stress-free cooling, this austenite texture transforms into four (8)-pole fiber texture of self-accommodated martensite. These four broad poles in AD IPF (Fig. 6a) consist of 8 maxima reflecting the lattice correspondence between <112> austenite direction and multiple martensite directions in variously oriented martensite variants in the self-accommodated martensite (for more information on fiber textures in NiTi wires see Ref. [31]).

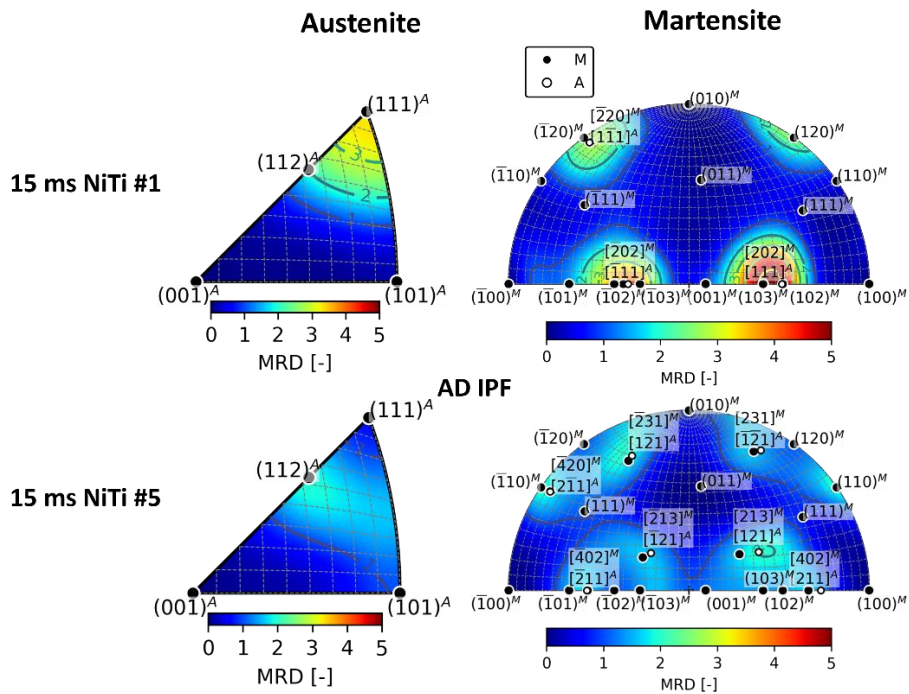

**Fig. 6b: Comparison of the fiber textures of austenite and self-accommodated martensite in superelastic NiTi#1 wire and NiTi#5 SME wire.** The austenitic texture of NiTi#5 SME wire is less sharp and displays broad maximum around the [112] pole. Based on this the fiber texture of martensite is also less sharp and displays four (8) maxima due to lattice correspondence [31]

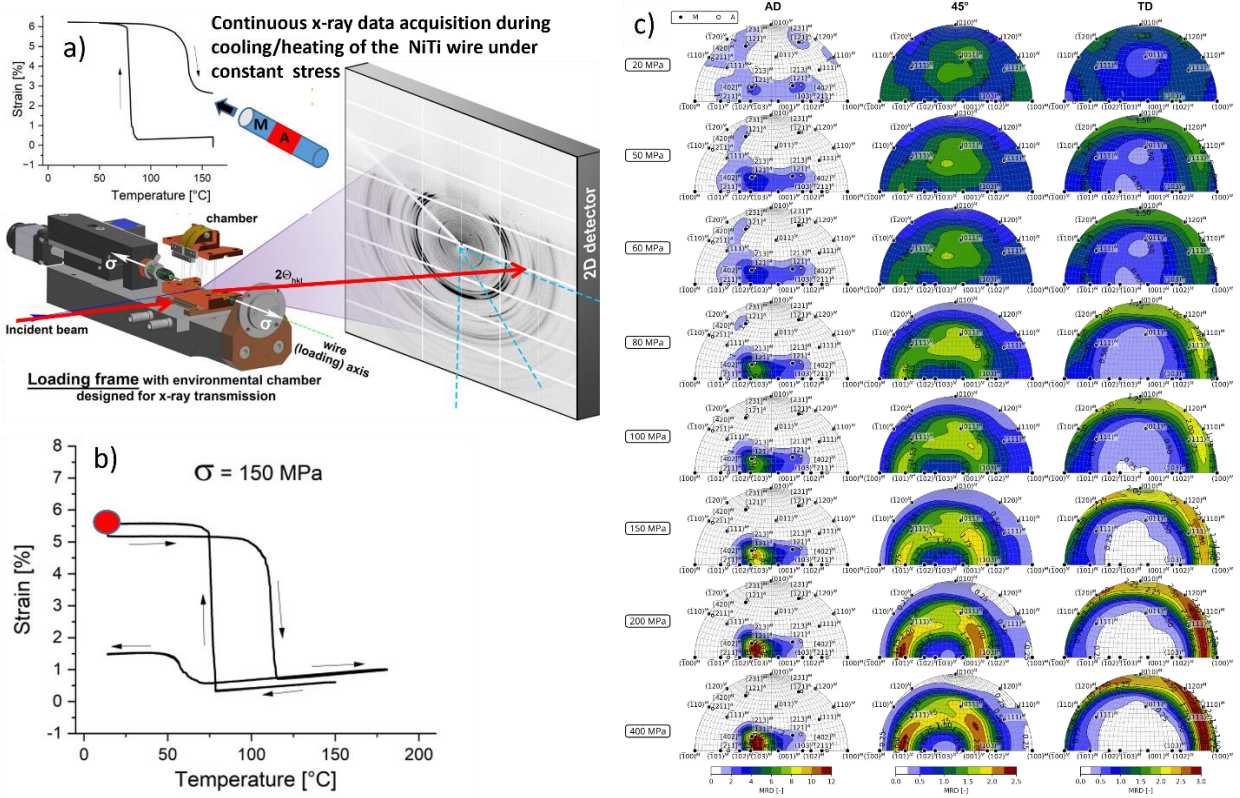

**Fig. 7: Textures of martensite created by cooling under stresses 0-400 MPa.** The textures evaluated at room temperature after cooling under stress (red spot in (b)) are presented in a form of inverse pole figures of axial, 45° and transverse directions to the wire axis (c).

When the forward MT proceeds under stress martensite textures depend on the magnitude of the applied stress due to preferential selection of martensite variants (Fig. 7c). Clearly, the texture of martensite becomes more and more oriented with increasing external stress in the 0-200 MPa stress range. Texture of oriented martensite is a single-pole [10-3] fiber texture with a tail towards the [213] direction on the right (for interpretation of IPFs of oriented martensite in terms of martensite variant selection see Ref. [31]). Selected results from thermomechanical loading tests involving cooling under 100 MPa and 400 MPa stress are shown in Figs. S7, S8.

### 3.3 Martensite variant microstructure created by forward MT upon cooling under external stress

Before presenting the reconstructed martensite variant microstructures in grains of deformed NiTi wires, let us first explain why we did that. Main reason is that the martensite variants created by the forward MT arrange themselves in martensite variant microstructures characteristic for the stress state under which they were created and/or for the deformation/transformation processes occurring afterwards. For example, self-accommodated microstructures are characteristic for MT taking place upon stress-free cooling [1,2,3,5,6], martensite variant microstructure observed in wires deformed up to the end of the transformation stress plateau in tensile test is characteristic for stress induced MT [7], microstructures observed in plastically deformed wires [32] are characteristic for the plastic deformation process, etc.

By reconstructing the martensite variant microstructures created in grains by the forward MT upon cooling under external stress, we want to learn how the recoverable transformation strains are generated as well as to reveal the mechanism by which the stress induced martensite deformed plastically. There is a general problem with TEM observation of microstructures in deformed SMAs which often focusses on dominant features without knowing whether they are relevant for the deformation/transformation mechanism that created them. Since individual martensite variants forming on nanoscale are differently oriented with respect to the electron beam, BF image of martensite variant microstructures does not provide meaningful information due to the peculiar electron diffraction contrast that is very sensitive to crystal orientation. Therefore, martensite variant microstructures in NiTi need to be analysed by nanoscale orientation mapping and/or by dark field imaging of the microstructure using carefully selected diffraction spots [30].

We assume that martensite variant microstructures observed within the disc shaped TEM lamella are the same as those existing in the wire under external stress at room temperature. Since the $M_f$ temperature of the 15 ms NiTi #1 wire is 63 °C, the wire cooled under external stress down to 20 °C and unloaded is fully martensitic and the assumption that the microstructure is retained on unloading is logical. Although we reconstruct the microstructure within the part of a grain within the lamella, we assume that same microstructure fills entire grain regardless of their size and shape [30].

Martensite variant microstructures need to be reconstructed within entire grains and in the same orientation of the martensite matrix with respect to the electron beam. In fact, the polycrystal grains (d=~250 nm) serve as a kind of "voxels", in which martensite variant microstructures are reconstructed. Surprisingly the reconstructed microstructures in all grains within the lamella are in some sense qualitatively similar regardless of the orientation of the austenite lattice with respect to the wire axis [30].

When reconstructing martensite variant microstructures in deformed NiTi wire by the SAED-DF method [7,30], the TEM lamella is tilted in such a way that the selected grain is oriented into [010] low index zone in martensite. In this zone, all crystal lattices within a single grain are equally oriented with respect to the electron beam and all interface planes are parallel to the electron beam [30]. Since the monoclinic martensite has low symmetry, there is only single [010] direction in the martensite lattice and, hence, it is not always possible to tilt the lamella so that the selected grain is oriented into the [010] zone. Martensite variant microstructure within grains of any orientation will be reconstructed also by the automated orientation mapping ASTAR method [30].

Fig. 8 provides an overview on martensite variant microstructures in whole grains of the NiTi wire cooled under 50,200,300,400, 500 MPa and 600 stresses. From top to bottom, individual figures show bright field (BF) images (1) of selected grains in general orientations that becomes all dark when oriented in the [010] zone (2) since all microstructure constituents within the grain diffract strongly. The composite electron diffraction patterns (3) with coloured reciprocal lattices (4) characterize the variously oriented martensite variants within the grain. Dark field (DF) images (5,6) provide information on spatial distribution of martensite variants within the grain.

The martensite variant microstructure observed in grains of the wire cooled under 50 MPa stress (below the reorientation stress 150 MPa) consists of multiple domains of mutually misoriented (001) compound twinned laminates ($\varepsilon^{tot}$ = 2% in Fig. 4b). This microstructure (Fig. 8) is similar to microstructures observed in self-accommodated martensite formed by stress-free cooling from the austenite (Fig. 1). The microstructures observed in the wire cooled under 200 MPa stress (just above the reorientation stress 150 MPa) contain single domain of (001) compound twins filling entire grains ($\varepsilon^{tot}$ = 6.3% in Fig. 4g). After cooling the wire under 300 MPa stress, the grains become partially detwinned and newly contain isolated (100) twins ($\varepsilon^{tot}$ = 7.5% in Fig. 4h). After cooling the wire under 400 MPa stress, the martensite within some grains becomes completely detwinned from (001) compound twins and there are (100) twin bands (single band on the right side of the grain in Fig. 8) ($\varepsilon^{tot}$ = 8.8% in Fig. 4i). After cooling the wire under 500 MPa stress, the detwinned microstructure contains multiple (100) twins and few (20-1) twins forming wedges (bottom part of the grain in Fig. 8), which evidences onset of plastic deformation of martensite by kwinking ($\varepsilon^{tot}$ = 11.4% in Fig. 4j). After cooling the wire under 600 MPa, the microstructure contains multiple (100) twins and multiple (20-1) kwink bands forming wedges which evidences massive plastic deformation of martensite by kwinking ($\varepsilon^{tot}$ = 15% in Fig. 4l).

It is important to point out that the observed martensite variant microstructures (Fig. 8) and martensite textures (Fig.7) are characteristic for the stresses under which the forward MT upon cooling occurred. Specifically, self-similar martensite variant microstructures were observed in any grain within the TEM lamella cut from the NiTi wire cooled under specific stress when it was oriented into the [010] zone. The self-similarity consists in common [010] zone shared by all microstructure constituents, in characteristic mutual misorientations among the diffracting blocks of martensite lattice and in orientation of all interfaces in the grain within the [010] zone. It is particularly the recorded composite electron diffraction patterns (rows 3,4 in Fig. 8) that are similar in all grains oriented into the [010] zone that represents this self-similarity. The differences among individual grains are in volume fractions of particularly oriented blocks of martensite lattice and in their spatial arrangement within grains.

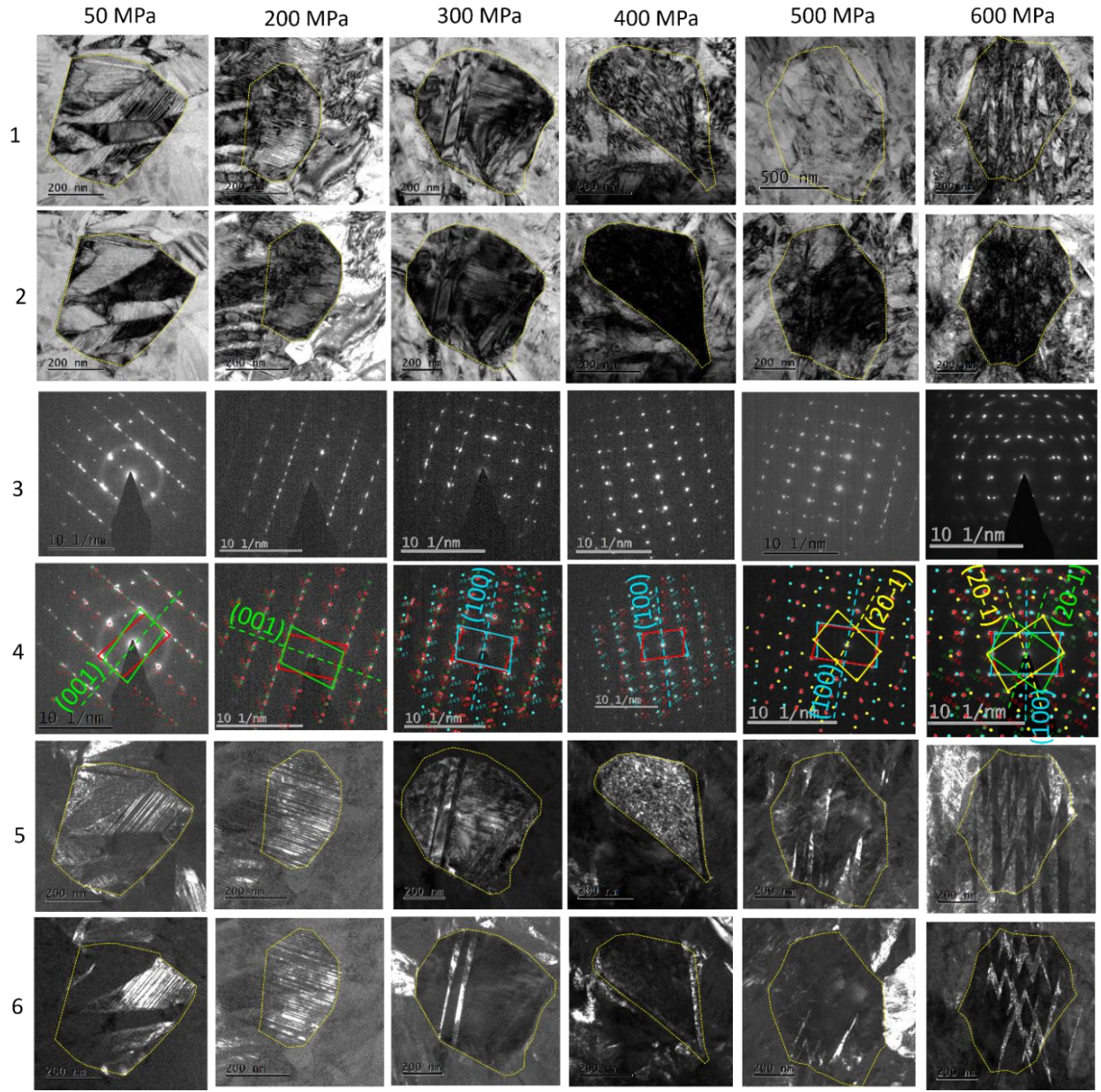

**Fig. 8: Martensite variant microstructures in 15 ms NiTi#5 wire created by forward MT upon cooling under stresses 50,200,300,400,500,600 MPa.** The successive rows show 1…BF image in general orientation, 2…BF image in [010] or [1-10] zone, 3… composite diffraction pattern, 4… composite diffraction pattern with coloured lattices of microstructure constituents, 5…DF image of martensite matrix, 6…DF image of martensite twins. In the case of 500 MPa and 600MPa stress, fifth row shows (100) twins and 6 row shows (20-1) twins (reflection used to take the DF image is common for both twins).

Below we present three examples of reconstructed martensite variant microstructures created by forward MT upon cooling under 50 MPa (Fig. 9), 300 MPa (Fig. 10) and 600 MPa (Fig. 11). The microstructures reconstructed by SAED-DF and ASTAR methods exist as ASTAR datasets– basically a 3D digital models of the martensite variant microstructure within the grain containing spatial, orientational and phase specific information on phases and martensite variants within the grain. We are presenting the reconstructed microstructures by showing three projections in x,y,z orthogonal directions ((l,m,n) in Figs. 9-11). Virtual bright field (VBF) and virtual dark field (VDF) images are generated from the ASTAR dataset.

The martensite variant microstructure in Fig. 9 is an example of microstructure formed upon cooling under very low stresses ($\varepsilon^{tot}$ = ~2% in Fig. 4b). There are multiple domains of (001) compound twinned laminate within the grain. As there is no common low index zone shared by the domains within the grain, it is not possible to tilt the lamella so that whole grain becomes dark (Fig. 9b). It is tilted such that some domains are oriented into the [010] zone but other domains are oriented generally. The recorded electron diffraction pattern (Figs. 9c,d,e) thus originates only from two dark domains (Fig. 9b), in which the (001) compound twin laminate can be visualized by DF imaging (Fig. 9f,g) since the twin planes are parallel to the electron beam. The reconstructed microstructure is visualized by the ASTAR orientation maps along x,y,z directions (Figs. 9l,m,n) coloured as suggested in the IPF (Fig. 9o). Other domains within the microstructure can be visualized by the VDF images (Figs. 9h,i,j,k). It is not very clear how this microstructure was created, it must have been from multiple martensite nuclei in which (001) compound twins formed. Although the wire was 2% elongated upon cooling (Fig. 4b), the martensite lattice in the grain is not suitably oriented for tension (the martensite displays shortening along the [010] crystal direction with respect to austenite (Fig. S9) which is, however, oriented along the wire axis (Fig. 9i)).

The martensite variant microstructure in Fig. 10 is an example of microstructure consisting of partially detwinned martensite matrix with two parallel (100) twin bands ($\varepsilon^{tot}$ = ~7.5% in Fig. 4b) in the reconstructed microstructure (Fig. 10l.m.n). The (100) twin bands are clearly visible in DF and VDF images (Figs. 10f-k). This microstructure, along with microstructures 200 MPa and 300 MPa in Fig. 8, are typical martensite variant microstructures of reoriented martensite in NiTi. Similar microstructures in grains were observed after tensile deformation in martensite beyond the reorientation plateau or after tensile deformation in austenite beyond the transformation stress plateau [7]. It is noteworthy, that polycrystal grains within the reoriented martensite (after cooling under 200-400MPa in Fig. 8) become effectively martensite single crystals (there is only single (001) crystal plane in each grain), except of the few isolated (100) twin, where the (001) plane is only slightly rotated (Fig. 9e). This is very important since such oriented martensite lattice in grains is prone to dislocation slip, as will be discussed in section 4.

The martensite variant microstructure in Fig. 11 is an example of the microstructure in a grain of the NiTi wire, in which the martensite deformed plastically by kwinking [33] ($\varepsilon^{tot}$ = ~15% in Fig. 4b). It contains (001) compound twinned matrix (Fig. 11h) with several, vertically oriented, parallel (100) deformation twin bands (Fig. 11g) and inclined (20-1) kwink bands (Fig. 11f). The matrix as well as the kwink bands remained (001) compound twinned. Microstructures in other grains (Fig. 8) frequently contained higher density of kwink bands. For example, the wedge microstructure in Fig. 8 is composed from (20-1) kwink bands in twinned orientation to the matrix and other (20-1) kwink bands in twinned orientation to the (100) twin (see DF image in Fig. 8). The matrix as well as the kwink bands in this microstructure are nearly detwinned. The readers interested in kwinked microstructures, variants and interfaces in plastically deformed B19' martensite are referred to dedicated articles [30,32,33]

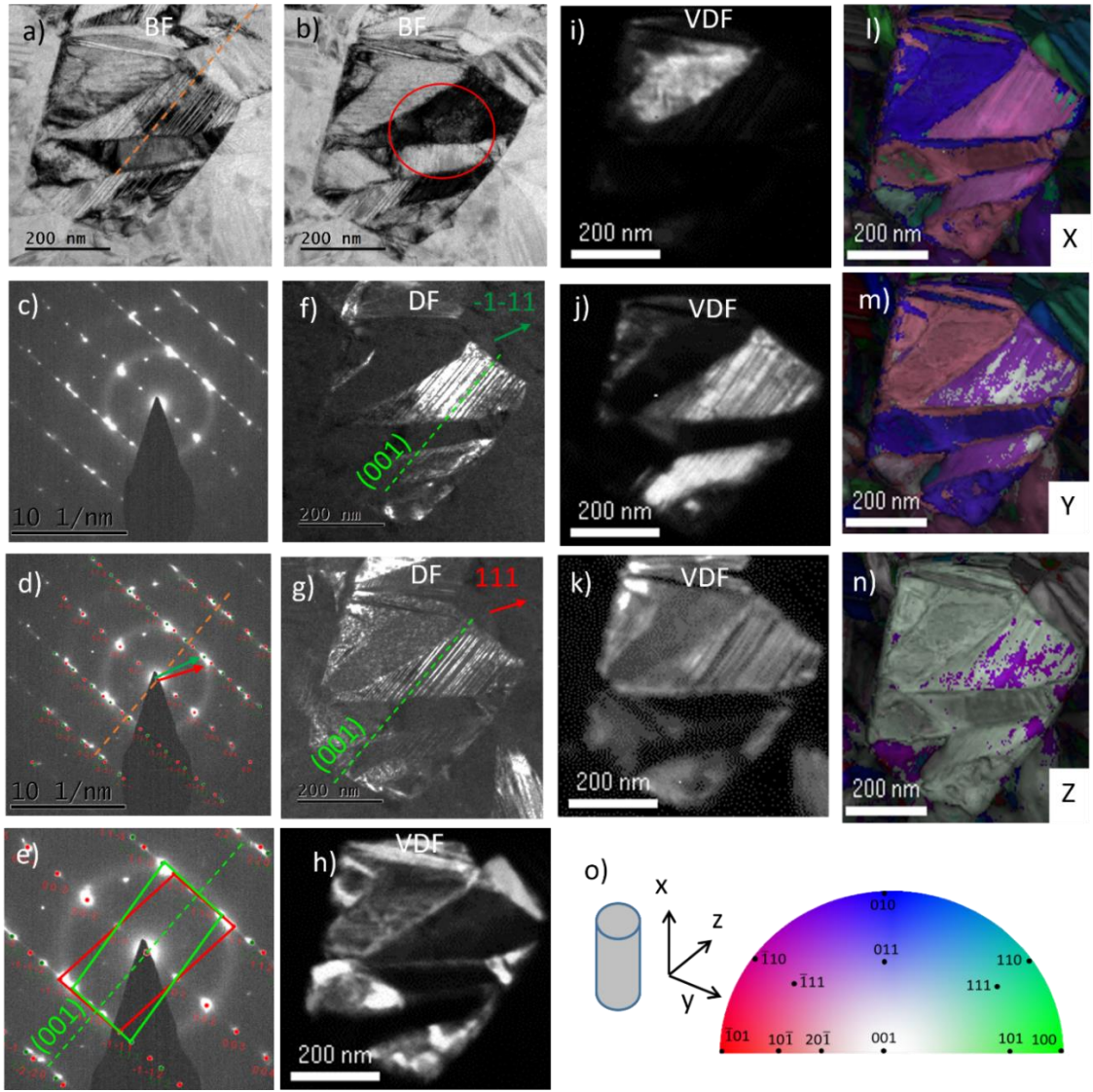

**Fig. 9: Martensite variant microstructure in 15 ms NiTi#5 wire created by forward MT upon cooling under 50 MPa stress** reconstructed by the ASTAR method. All grains contain multiple martensite variants. When a domain containing thin laminates is oriented in [1-10] low index zone (b), other domains within the grain remain bright. The composite diffraction pattern (c) originates mainly from the dark domains in (b), but there are additional spots (d,e). DF images show the matrix (f), (001) twins (g) within the dark domain visualize the (001) compound twin laminates in domains. Other domains are not indexed in (e). This information is a part of the ASTAR reconstruction (l,m,n) of the grain. As the (001) twin planes are inclined to the electron beam, they cannot be resolved in figures (l,m,n). However, they are clearly resolved in DF images (f,g) and virtual dark field (VDF) images (j.k). The coordinate system with the z-axis along the wire axis and colored orientation space of the monoclinic lattice (o) help to understand the reconstructed microstructure.

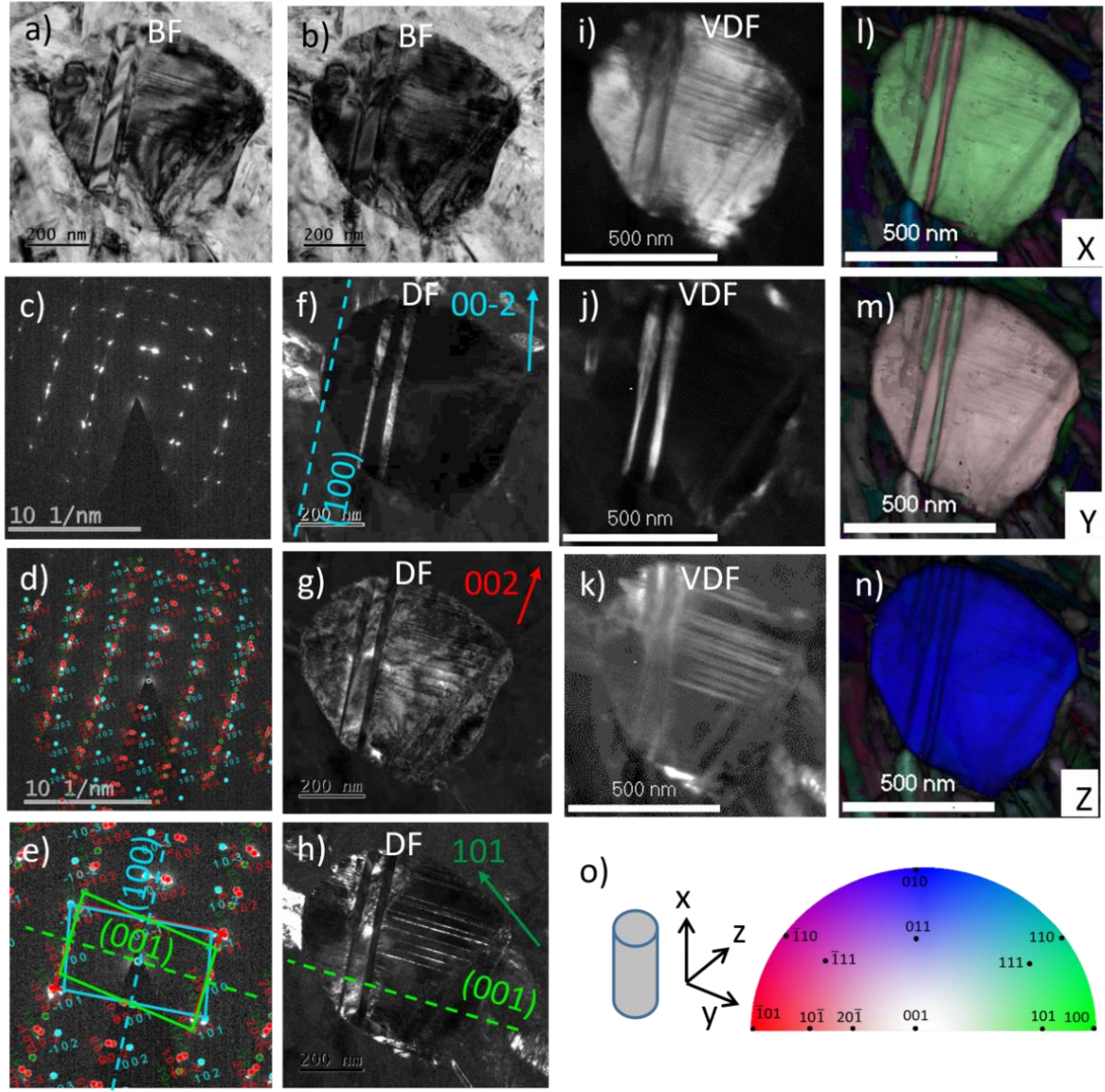

**Fig. 10: Martensite variant microstructure in 15 ms NiTi#5 wire created by forward MT upon cooling under 300 MPa stress** reconstructed by the ASTAR method. a) selected grain in general orientation. When the TEM lamella is tilted so that selected grain is oriented in [010] low index zone, entire grain becomes dark (b). The composite electron diffraction pattern recorded in this orientation (c,d,e) corresponds to three martensite lattices misoriented in the [010] zone (d,e). DF images (f,g,h) show bright martensite matrix (g), (001) twins (h), (100) twins (f). ASTAR reconstruction (l.m.n) cannot resolve the (001) compound twin laminates. They are clearly resolved in VDF images (i,j,k). The coordinate system with the z-axis along the wire axis and coloured orientation space of the monoclinic lattice (o) are needed to understand the reconstructed microstructure.

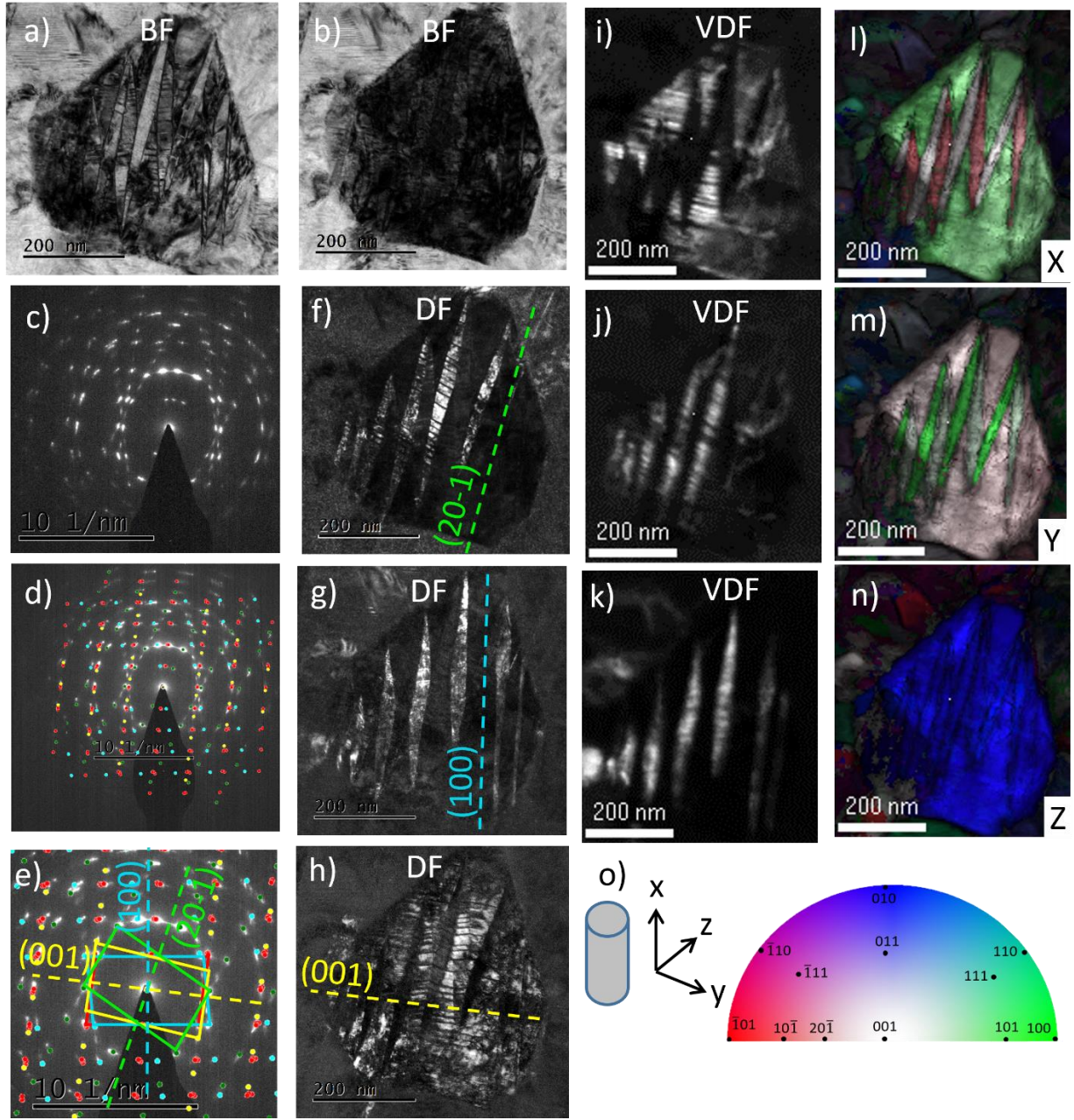

**Fig. 11: Martensite variant microstructure in 15 ms NiTi#5 wire created by forward MT upon cooling under 600 MPa stress** reconstructed by the ASTAR method. a) selected grain in general orientation. b) selected grain oriented in the [010] low index zone. The composite diffraction pattern (c,d,e) corresponds to three martensite lattices misoriented in the [010] zone (d,e). DF images show bright martensite matrix (h), (20-1) twins (f), (100) twins (g). Both matrix and twins (100) (20-1) contain laminate of (001) compound twins (h). The ASTAR reconstruction (l,m,n) cannot resolve the (001) compound twin laminates. They are, however, clearly resolved in the VDF images (i,j,k). The coordinate system with the z-axis along the wire axis and coloured orientation space of the monoclinic lattice (o) are needed to understand the reconstructed microstructure.

**3.3 Dislocations defects in martensite created by forward MT upon cooling under external stress**

Since the forward MT upon cooling under stress generates plastic strains and since we assumed that plastic deformation occurred via dislocation slip in martensite, we searched for slip dislocations in martensite variant microstructures in NiTi wires cooled under 200, 300 and 400 MPa stresses (Figs. 8,10). We tried to characterize the observed dislocation defects by conventional TEM. However, this turned out to be very difficult due to the peculiar electron diffraction contrast and due to presence of (001) compound twins in martensite. Therefore, we applied geometric phase analysis (GPA) in HRTEM mode to detect slip dislocations, but as there were only very few slip dislocations in analyzed grains, we have not succeeded.

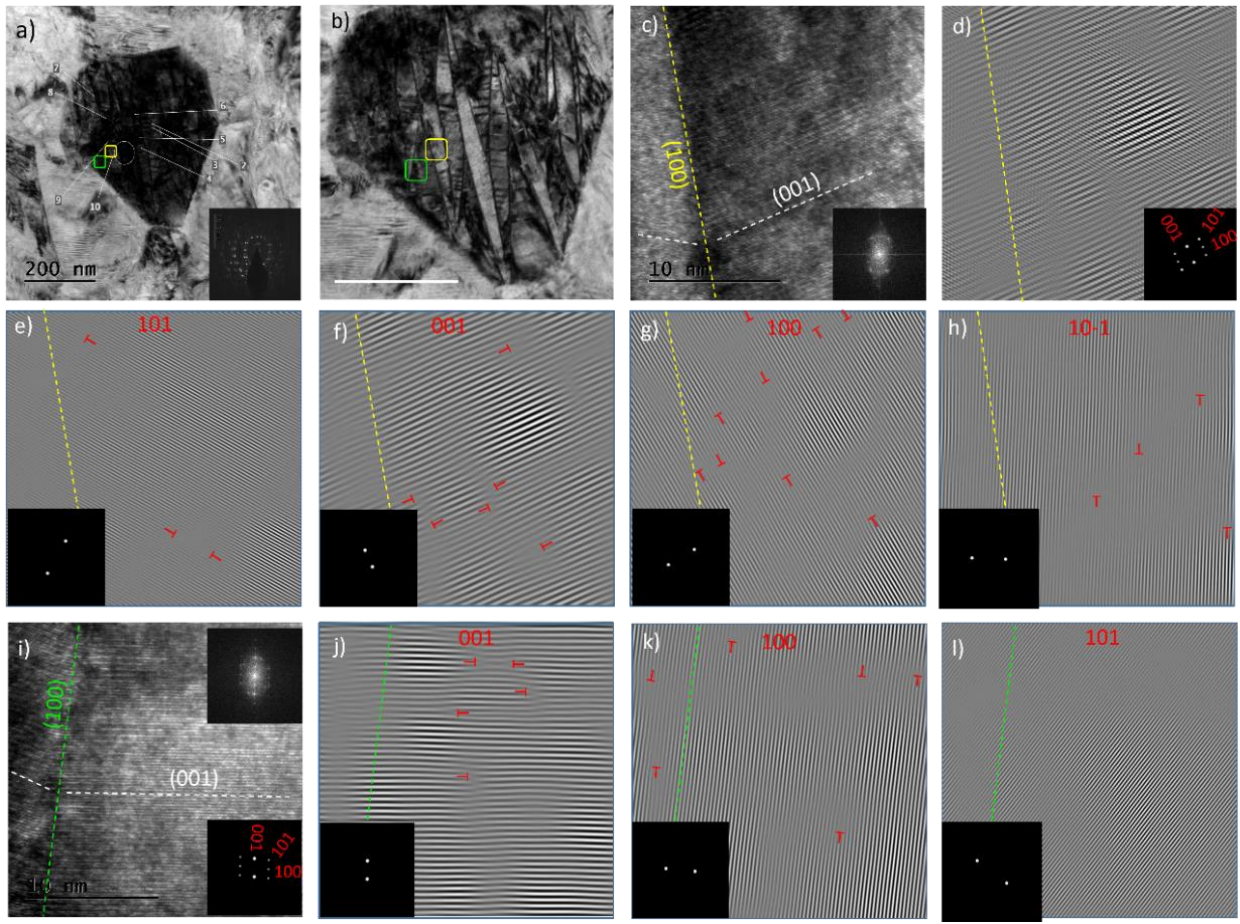

**Fig. 12:** Slip dislocations in martensite plastically deformed by kwinking in the microstructure of the NiTi wire cooled under 600 MPa stress. Selected location within the grains oriented in the [010] zone (yellow and green square in (a,b)) was observed in HRTEM mode (c,i). GPA analysis of dislocation defects was performed using 4 reflections (d-h, j-l). There are (100) twin interfaces denoted by yellow and green dashed lines. The observed slip dislocations are marked.

Therefore, we analyzed slip dislocations in the microstructure of the NiTi wire cooled under 600 MPa which displays plastically deformed martensite with multiple kwink bands (see selected grain in Fig. 11). The GPA analysis showed presence of slip dislocations with Burgers vector [100] (Fig. 12g) in any HRTEM image taken within the analyzed grain (Fig. 12). Slip dislocations with other Burgers vectors (Fig. 12d,f,h) were less frequent. The frequent observation of slip dislocations with Burgers vector [100] on (001) crystal plane is understandable, since activation of kwinking deformation involves coordinated slip of [100](001) slip dislocations in martensite [33].

## 4. Discussion

**4.1 Martensite variant microstructures and textures created the forward MT upon cooling under external stress**

The experiments (Figs. 4,5) have shown that maximum recoverable strains are achieved upon cooling the NiTi wire under surprisingly low external stress 200 MPa, only 50 MPa above the reorientation stress (Fig. 3). This result is surprising, because researchers relying on the results of standard thermal tests (note the strain $\varepsilon^{tot}$ increasing with increasing external stress in Figs 4,5) would not expect this result. Nevertheless, it is logical and it is in accord with the observation of martensite variant microstructures (Fig. 8-11) and martensite textures (Fig. 7) created by the forward MTs upon cooling under stress, which both change with stress increasing from 0 to 200 MPa but remain unchanged with further increasing stress..

When the external stress applied on cooling exceeds the reorientation stress 150 MPa, most of the grains observed in the TEM lamella contained martensite variant microstructure consisting of single laminate of (001) compound twins and martensite texture characteristic for oriented martensite (Figs. 7,8,10). Since we observed similar microstructures in grains of the NiTi wire deformed in martensite state beyond the end of reorientation plateau up to 400 MPa [7], we were not surprised to see it after cooling under external stress.

One may wonder why there are laminates of (001) compound twins in the microstructure instead of laminates of type II twins and why the (001) compound twins do not detwin under stress exceeding the reorientation stress (it would add additional ~5% macroscopic strain). Our explanation is that the (001) compound twins were introduced by the habit plane of the forward MT in small grains, as will be discussed in the next section 4.2. It needs to be mentioned here that (001) compound twins were observed in thermally as well as stress induced martensite in nanocrystalline NiTi alloys by many investigators [6,7,14].

As concerns the question why detwinning of (001) compound twins did not occur during the cooling under constant stress (Fig. 8), we assume it originates from the requirement for strain compatibility at grain boundaries in SMA polycrystals, as elaborated and discussed by Bhattacharya and Kohn [45]. This constraint on the detwinning of (001) compound twins stemming from grain interactions was specified for the particular case of <111> fiber textured NiTi polycrystal in our earlier work (Appendix C in Ref. [7]).

As the external stress applied on cooling increased up to 300-500 MPa, partially detwinned single domain martensite variant microstructures were frequently observed in grains (Figs. 8,10). Simultaneously (100) twin bands appeared in some grains and their density increased with increasing applied stress. The appearance of (100) twins in the microstructure is an apparent puzzle, since (100) twinning of oriented martensite matrix does not increase the tensile strain of the oriented martensite matrix (see Fig. A4b in [31]). The (100) twinning becomes activated upon cooling under ~400 MPa stress because it opens the way for plastic deformation of martensite by kwinking [32,33].

Kwinking deformation takes place upon cooling under external stresses that are significantly lower than the kwinking stress determined in isothermal tensile tests [7,32,33]. With increasing external stress, at first (100) deformation twinning is activated at 300 MPa (Fig. 8), then (20-1) kwink bands appear at 500 MPa (Figs. 8,10) and, finally, the wedge microstructure created by the forward MT appear in the martensite variant microstructures in grains of the wire cooled under 600 MPa stress (Figs. 8,11). By reconstructing the martensite variant microstructures in NiTi wires cooled under stresses 300-600 MPa, we were able to get new insight into subsequent stages of kwinking deformation that was no possible to obtain from the analysis of kwinked microstructures in NiTi wires deformed in isothermal tensile tests [7,32,33]. The wedge microstructures in NiTi wire cooled under 600 MPa stress (Figs. 8,11) are very similar to martensite variant microstructures created by plastic deformation of martensite by kwinking in isothermal tensile tests.

**4.2 Habit plane of the forward MT upon cooling under external stress**

Since martensite variant microstructures observed in grains of thermally induced self-accommodated martensite consist of multiple domains of (001) compound twin laminate (Fig. 1), we assume that forward martensitic transformation taking place upon stress free cooling of the 15 ms NiTi #5 wire proceeds via a habit plane interfaces between austenite and second order (001) compound twin laminate of martensite (0 MPa in Fig. 13), as proposed by Waitz [6]. The self-accommodated martensite is characterized by four (8)-pole AD IPF fiber texture (Fig. 7) and neither plastic strains nor lattice defects are generated by stress-free thermal cycling across transformation interval [10]. This habit plane regime is considered to apply for both forward and reverse martensitic transformations proceeding in the absence of stress regardless of the thermomechanical loading history - i.e. reverse martensitic transformation of self-accommodated, oriented or plastically deformed martensite is assumed to take place via the habit plane between austenite and second order (001) compound twin laminate of martensite and no plastic strains and lattice defects are assumed to be generated. It needs to be mentioned that the sketch in upper left corner of Fig. 13 is only a simplification enabling the reader to understand why (001) compound twins appear in martensite variant microstructures created by the forward MT upon cooling, we have no aspiration to discuss crystallographic details of these habit plane interfaces. In contrast to nanograin microstructures (d~ 20 nm) analyzed by Waitz [6], multiple variously oriented (001) compound twin domains appear in ~250 nm grains of the 15 ms NiTi #5 wire (Fig. 1) suggesting that the stress-free MT occurred via propagation of multiple habit planes in each grain.

When the forward MT takes place under low stress (100 MPa in Fig. 13), it proceeds via the same habit plane, but there is a difference in the martensite variant microstructure it created. Many grains contain (001) compound twin laminates oriented in [010] zone (50 MPa in Fig. 8, Fig. 9). We explain this by assuming that the stress induced martensite immediately starts to reorient, but the reorientation is not completed since the external stress is too low (Fig. 9). No plastic strains and no lattice defects are generated by such forward MT proceeding under low stresses below the reorientation stress.

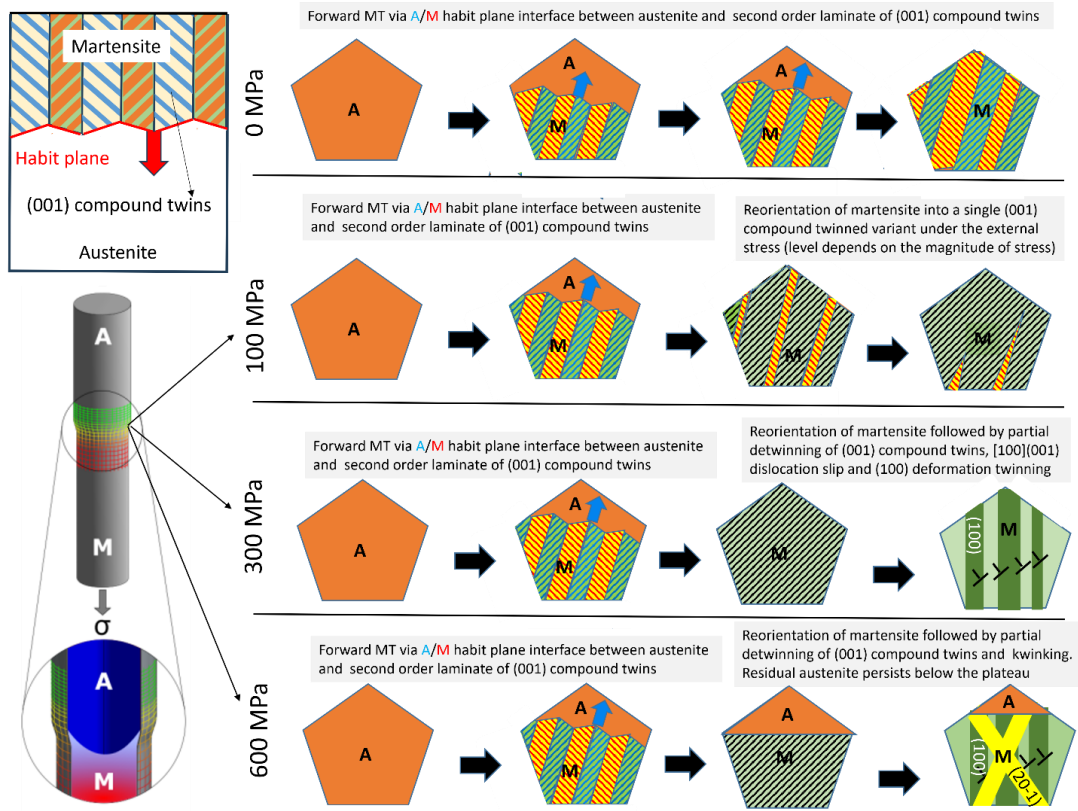

**Fig. 13: Habit plane of the forward martensitic transformation upon cooling under various applied stresses.** With or without the applied stress, the forward MT proceeds via strain compatible habit plane interfaces into second order laminate of (001) compound twinned martensite. Depending on the magnitude of the applied stress, the newly created martensite immediately reorients into single domain (001) compound twin laminates filling in each grain and deforms plastically by dislocation slip in martensite (and by kwinking at highest stresses). All suggested processes occur within macroscopic interface propagating along the wire during the forward MT.

When the forward MT on cooling takes place under medium stresses (300 MPa in Fig. 13), it proceeds via the same habit plane as under low external stresses, but the martensite immediately reorients and partially detwins. In addition, many grains contained (100) twins (Figs. 8,10), the amount of which increased with increasing stress. The texture of the stress induced martensite dominated by single (10-3) pole corresponding to the reoriented martensite [31]. A question remains how the experimentally observed plastic strain (e.g. ~3% under 400 MPa stress in Fig. 5) was generated by the forward MT under external stress (Fig. 5). The plastic deformation of martensite by kwinking did not start yet and plastic deformation via [100](001) dislocation slip is constrained (Appendix C in [7]). We explain this by considering that the [100](001) dislocation slip in martensite partially occurs in the 15 ms NiTi #5 wire, in spite of the polycrystalline constraint (Fig. S5), as will be discussed in section 4.4.

When the forward MT on cooling takes place under largest stresses (600 MPa in Fig. 13), it is still considered to proceed via the same habit plane (as (001) twins are still observed within the microstructure (Fig. 11)), but the martensite promptly reorients, detwins and deforms plastically via dislocation slip and kwinking in martensite (see kwink bands within the microstructure in Figs. 8,11). Since the wire continues to deform upon cooling under high stresses 500-600 MPa even below the end of the temperature plateau (Fig. 4j.k.l), we assume that the forward MT is not completed at the end of the temperature plateau upon cooling under stress and small volume fraction of austenite in the wire transforms to martensite upon further cooling to the room temperature. The most likely explanation of this retained austenite is based on the assumption of internal stresses arising due to incompatible shape strains of single martensite variants in grains created within the temperature plateau. The martensitic transformation of the retained austenite upon further cooling under stress is accompanied by plastic deformation of martensite via dislocation slip and kwinking that allows to overcome the strain incompatibilities arising at grain boundaries (similar scenario was proposed for tensile deformation of the same wire beyond the end of the transformation stress plateau in isothermal tensile test [14]).

**4.3 Permanent lattice defects created by the forward MT upon cooling under external stress**

Only few dislocations were found in the microstructure of the 15 ms NiTi #5 wire cooled under 300 MPa. On the other hand, slip dislocations and kwink bands were systematically observed in the microstructure of the wire cooled under 600 MPa (Fig. 12). This was explained by plastic deformation of the martensite by kwinking which proceeds via coordinated dislocation slip. The problem is lack of slip dislocations observed in NiTi wires cooled under medium stresses. It is not clear how the 2% plastic strains (Figs. 4,5) were generated upon cooling under 300 MPa stress.

Another way how to observe permanent lattice defects created by forward MT upon cooling under stress is to analyze dislocation defects in austenite after stress free heating above the $A_f$ temperature. Since this is difficult with the 15 NiTi#5 SME wire, we performed such observation on TEM lamella cut from 15 ms NiTi #1 superelastic wire [17] and selected results are shown in Fig. 14. There were no slip dislocations in 15 ms NiTi #1 wire cooled under 400 MPa stress [17]. This evidences that this NiTi wire can undergo forward MT under stress significantly higher than the reorientation stress without generating plastic strains and dislocation defects.

Slip dislocations belonging to <100>{011} slip systems were observed in the NiTi #1 wire cooled under 700 MPa (Fig. 14). The observed slip dislocations are heterogeneously distributed both among and within austenite grains. Many authors observed similar slip dislocations in superelastically cycled NiTi before [12,15,46], but no one has ever explained satisfactorily their origin. A problem is that these dislocations were observed in the literature in austenite phase on TEM foils cut from the NiTi subjected to closed loop superelastic cycle, which means that it could not be specified whether they were created by the forward or reverse MT. In a contrast to earlier works [12,15,46], slip dislocations in Fig. 14 were generated solely by single forward MT most likely by dislocation slip in martensite. Key result of the work [17] is that the forward MT generates different dislocation defects than the reverse MT, not only as concerns their density but also their character.

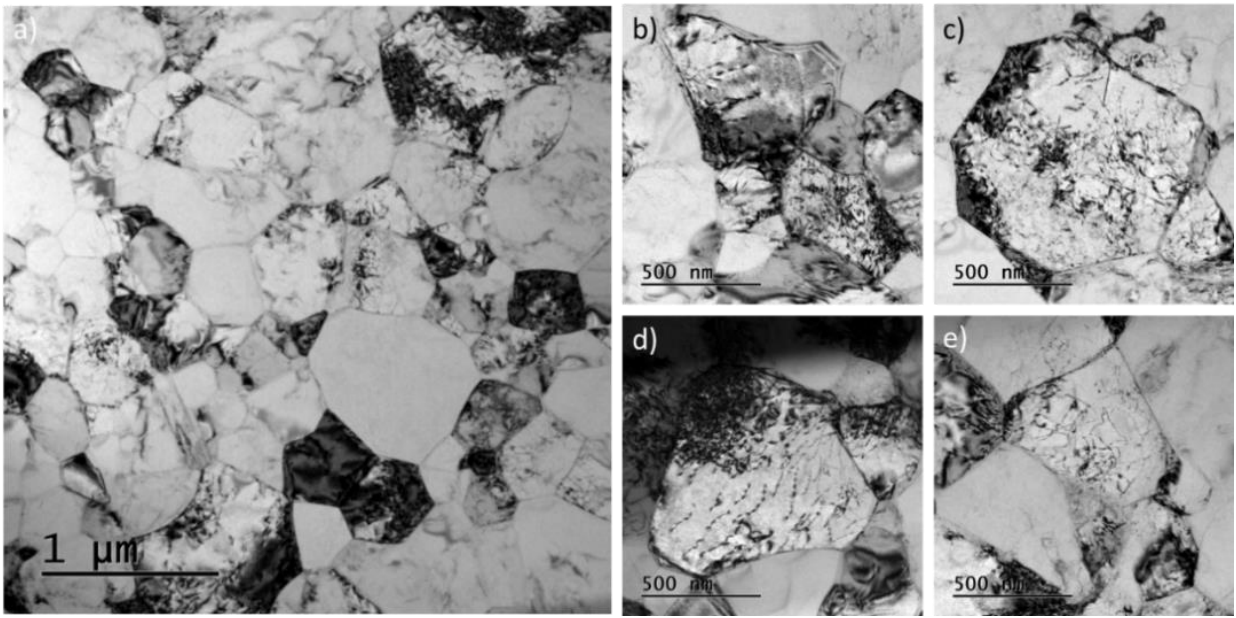

**Fig. 14: Dislocations observed in 15ms NiTi #1 superelastic wire after forward MT upon cooling under 700 MPa, unloading and stress-free heating to room temperature** [17]. Bright field TEM images in two magnifications. The forward MT in in 15ms NiTi #1 wire generated ~2 % plastic strain, which equals to plastic strain generated by forward MT upon cooling the 15ms NiTi #5 wire under 300 MPa stress. Therefore, the slip dislocations shown in (b-e) shall be looked at as dislocations in the 15ms NiTi #5 wire cooled under 300 MPa stress. Since the TEM lamella was cut perpendicularly to the wire axis, most of the grains are oriented near <111> austenite zone and slip dislocations in {011} slip planes could be easily analyzed. See Ref. [17] for details lattice defects in austenite.

High density of slip dislocations and {114} austenite twins were observed in the superelastic NiTi wire cooled under highest stress 800 MPa, unloaded and heated to room temperature [17]. Since {114} austenite twins are known to be relicts of plastic deformation of martensite [7, 47,48], it can be deduced that the wire deformed plastically upon cooling at 800 MPa stress ($\sigma^Y_{M\_kwink}$ kwinking line meet the $\sigma^{FOR}$ forward transformation line in the σ-T diagram of the NiTi #1 wire at 900 MPa [11]).

**4.4 Recoverable and plastic strains generated by the forward MT upon cooling under external stress**

The 15 ms NiTi #5 SME wires cooled under stresses higher than 100 MPa contain grains filled with single domain (001) compound twin laminates (Fig. 8). Since the detwinning of (001) compound twins is constrained in polycrystalline environment (Appendix C in [7]), the maximum recoverable transformation strain is limited to ~5% (about half of the maximum recoverable transformation strain ~11% achievable from <111> fiber textured nanocrystalline NiTi wires). The experiments, however, also showed that the 15 ms NiTi #5 wire tends to deform plastically when exposed to stresses above 100 MPa (Figs. 5, S2). The plastic strains generated by tensile deformation in martensite state are relatively small and increase proportionally to the increasing recoverable transformation strain (Figs. S4,S5,S6). This is a very unusual mode of plastic deformation.

We assume that the small plastic strains generated by forward MT cooling under low stresses <400MPa (Fig. 5) are generated via [100](001) dislocation slip in martensite [35,32,33] equally as the plastic strains generated by tensile deformation in martensite state (Figs. S4,S5).. Since the Burgers vector/slip plane of the [100](001) dislocation slip are parallel to twinning shear direction/twinning plane of the (001) compound twining, both slip and twinning deformation processes are constrained in the polycrystalline environment (Appendix C in [7]). Since plastic strains generated by the forward MT under stresses increase with increasing stress, but the recoverable transformation strains remain constant (Fig. 5), we deduce that the (001) compound twinned martensite does not detwin but deforms via dislocation slip instead.

In order to explain why the B19' martensite in the 15 ms NiTi #5 SME wire deforms via the [100](001) dislocation slip even at the low stress (Fig. S2), we need to delve into crystallography of the B2-B19' martensitic transformation in NiTi. Fig. 15 visualizes atomic structures of B2 austenite and B19' martensite as "cells" built by the software CrysTBox for automated processing of TEM images and crystal structure visualization [49]. Looking at 2D projections of both crystal lattices perpendicularly to the (001)A and (-110)A austenite planes and lattice correspondent (100)M and (010)M martensite planes, the characteristic monoclinic distortion of martensite is clearly visible. The lattice can be distorted leftwards or rightwards giving rise to martensite variants V1 and V2. The weak Ti-Ti bonds of the monoclinic structure (indicated by double arrows in Fig. 15), enable easy shearing of the monoclinic lattice on (001) plane in the [100] direction [50], which facilitates both (001) compound twinning between the martensite variants V1 and V2 as well as dislocation slip on the (001) plane in the [100] direction. The (001) plane, as the single most densely packed lattice plane of the B19' monoclinic structure with largest lattice spacing, is simultaneously most preferable crystal plane for dislocation slip [44]. The oriented martensite thus can deform at very low stress by the [100](001) dislocation slip as well as by (001) compound twinning. Although both deformation mechanisms are constrained in polycrystalline constraint (Appendix C in [7]), small plastic strain proportional to recoverable transformation strains are recorded (Fig. 5,S4,S5,S6). While the (001) compound twin laminates yields crystallographically limited but recoverable strain, the [100](001) dislocation slip is capable of yielding unlimited plastic strains.

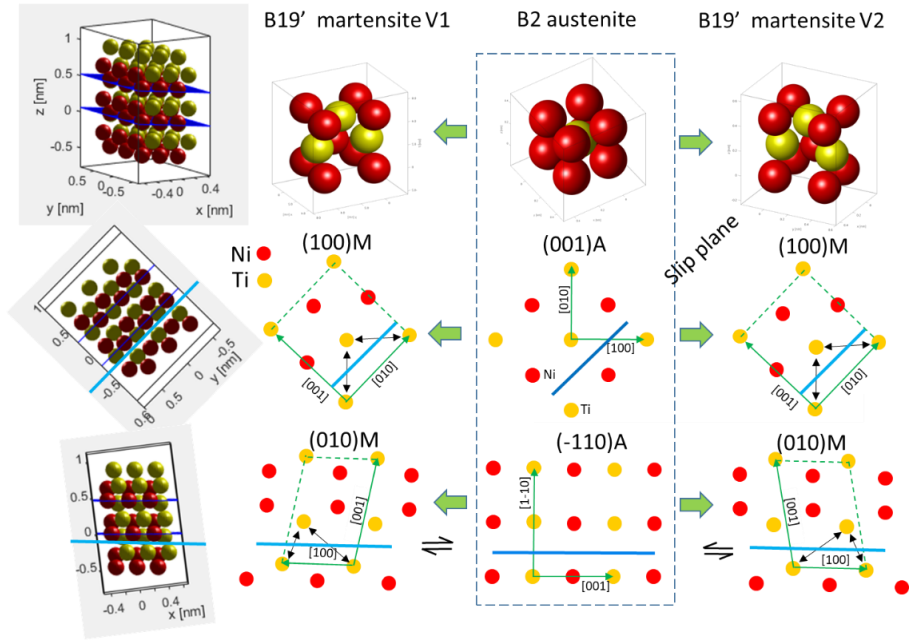

**Fig. 15: Dislocation slip and (001) compound twinning in monoclinic B19' martensite** The sketch shows how the crystal structure change during the B2-B19' martensitic transformation by showing atom positions on (001)A and (110)A cross sections along austenite planes changing into their positions on (100)M and (010)M cross sections along martensite planes, respectively. Two layers of atoms are plot together without differentiating between them. There are week Ti-Ti bonds (denoted by double black arrows) allowing for easy shearing the monoclinic structure [50]. Shearing in the [100] direction enables (001) compound twinning (lattice correspondence variants V1 and V2 can be converted one into another by the shear applied on (001) plane along the +/- [100] shear direction) and dislocation slip in [100](001) direction. The [100](001) and [010](001) martensite slip systems are inherited into austenite as <100>{110} and <110>{110} slip systems, respectively.

On the other hand, as already mentioned above, not all NiTi wires display plastic strains when deformed in tensile tests in the martensite state. The superelastic 15 ms NiTi #1 wire, when deformed in martensite state up to 900 MPa (12% strain), unloaded and heated in closed loop thermomechanical cycle (Fig. S1a,b), displays 9% fully recoverable transformation strain. This means that partial detwinning of (001) compound twins upon tensile deformation in martensite state becomes possible, if the stress is high enough and if the martensite can resist plastic deformation via dislocation slip, in spite of the constraint from neighboring grains (Appendix C in [7]). If the 15 ms NiTi #1 superelastic wire is deformed in martensite state further beyond the yield point up to 1100 MPa (14.5% strain), unloaded and heated in closed loop thermomechanical cycle (Fig. S1c,d), it displays also ~9% recoverable transformation strain but accompanied by ~1% plastic strain. Based on this we deduce that the 15 ms NiTi #1 superelastic wire is indeed more resistant to [100](001) dislocation slip in martensite than the 15 ms NiTi #5 SME wire.

Finally, let us point out again that the isothermal tensile loading of NiTi in the martensite state up to ~1GPa gives rise to small plastic strains ~2%, while the forward MT cooling under 600 MPa stress gives rise to very large plastic strains 8 %. Similarly, forward stress induced MT in isothermal tensile test generates 8% plastic strains (Fig. 7b in [14]). Although this is very well known in the SMA field (it is the reason why NiTi SME elements, if they are not to be damaged, have to be always deformed at low temperatures below Mf), the reason for it was not very clear, at least as far as we know. The reason is that the plastic deformation of oriented martensite via kwinking proceeding alongside the forward MT is not constrained by the polycrystalline constraint (Appendix C in [7]) and becomes activated at significantly lower external stress (Figs. 3,5, 8) than during the isothermal tensile loading in the martensite state (Figs. S5,S6). The reason why this is the case is linked to the coupled martensitic transformation with plastic deformation in NiTi which remains to be further investigated.

## 5. Conclusions

Martensite variant microstructures, martensite textures and plastic strains generated by forward B2-B19' martensitic transformation in nanocrystalline 15 ms NiTi #5 SME wire subjected to cooling under various external tensile stresses were investigated by thermomechanical testing, reconstruction of martensite variant microstructures in whole grains by TEM, textures in austenite and martensite phases by in-situ synchrotron x-ray diffraction and HRTEM analysis of permanent lattice defects in martensite.

The forward martensitic transformation upon cooling the NiTi wire under external stress:

1. proceeds via strain compatible habit plane interfaces into second order laminate of (001) compound twinned martensite. The newly created martensite reorients into a single domain (001) compound twin laminate filling in each grain which deforms plastically.
2. creates martensite variant microstructures in grains and martensite textures varying with increasing applied stress from multi domain microstructure of self-accommodated martensite to single domain (001) compound twin laminate filling entire grains of reoriented martensite at intermediate stresses up to complex wedge microstructures created by plastic deformation of martensite by kwinking upon cooling under highest stresses
3. generates recoverable transformation strains increasing with increasing external stress up to ~5% strain at ~200 MPa and remain constant with stress increasing further up to 600 MPa. The maximum recoverable transformation strain ~5% is about a half of the maximum recoverable transformation strain ~11 % achievable from nanocrystalline NiTi wires due to the (001) compound twinned microstructure in grains, less sharp <111> austenite fiber texture of the wire and possibly due to the [100](001) dislocation slip in martensite, which substitutes detwinning of (001) compound twins.
4. generates plastic strains via [100](001) dislocation slip in martensite from ~150 MPa stress and via plastic deformation of martensite by kwinking from ~500 MPa external stress.

The reconstruction of martensite variant microstructures in whole grains and martensite textures formed by the forward martensitic transformation upon cooling NiTi wire under various external stresses bring first ever experimental evidence on the martensite variant microstructures and textures in grains of NiTi polycrystal cooled under various external stresses.

## Acknowledgment

Support from Czech Science Foundation (CSF) projects 22-15763S (Heller) and 22-20181S (Sittner) is acknowledged. P. Šittner acknowledges support from Czech Academy of Sciences through Praemium Academiae. ESRF is acknowledged for support of the experiment MA5425. MEYS of the Czech Republic is acknowledged for the support of infrastructure projects, CNL (CzechNanoLab LM2018110) and Ferrmion (CZ.02.01.01/00/22_008/0004591).

## Data availability

Data will be made available on request.

ASTAR datasets with reconstructed martensite variant microstructures in Figs. 9,10,11 are available for downloading at http://ofm.fzu.cz/downloads